第一原理電子状態計算の新規手法に関する一提案：

記号的-数値的計算法

A proposal to first principles electronic structure calculation:

Symbolic-Numeric method

菊地章仁

Akihito Kikuchi

CANON INC.

30-2, Shimomaruko 3-chome, Ohta-Ku, Tokyo 146-8501-Japan

E-mail:kikuchi.akihito@canon.co.jp

**要約**

本研究は，記号的―数値的計算手法を利用した第一原理電子状態計算を試みる。ハートリー・フォック・ロータン方程式を用いた分子軌道計算において，基礎方程式を連立多項式方程式系と見なして近似を行った後，数式処理によって同一の零点を持つ連立多項式方程式系を作成，その解を数値的に求めることによって自己無撞着電子状態計算を実行する。こうすることで，電子状態計算．・最適化計算・逆問題計算を統一的に，すべて順問題として解くことができる。その際，試行錯誤的な反復計算は不要である。

**Abstract**

This study proposes an approach toward the first principles electronic structure calculation with the aid of symbolic-numeric solving. The symbolic computation enables us to express the Hartree-Fock-Roothaan equation in an analytic form and approximate it as a set of polynomial equations. By use of the Gröbner bases technique, the polynomial equations are transformed into other ones which have identical roots. The converted equations take more convenient forms which will simplify numerical procedures, from which we can derive necessary physical properties in order, in an "a la carte" way. This method enables us to solve the electronic structure calculation, the optimization of any kind, or the inverse problem as a forward problem in a unified way, in which there is no need for iterative self-consistent procedures with trials and errors.

## 1.背景

　第一原理的電子状態計算の基礎方程式は，電子－原子核系のエネルギー汎関数の極小条件に基づいて導出されるハートリー・フォック方程式あるいはコーン・シャム方程式である。これらの方程式はシュレーディンガー方程式の型に表される[1-3]。

$$\left(-\frac{1}{2}\Delta + \sum_a \frac{Z_a}{|r-R_a|} + \int dr' \frac{\rho(r')}{|r-r'|} + V^{exc}\right)\phi_i(r) = E_i\phi_i(r) \quad (1)$$

　左辺括弧内の第二項は，電荷 $Z_a$ をもつ原子核のポテンシャルである。第三項は，電荷分布 $\rho$ が作るクーロンポテンシャルである。第四項は，多電子系に働く量子力学的な相互作

用であり，交換相関ポテンシャルと呼ばれる。この方程式は，波動関数を何らかの基底関数系によって展開することによって数値的に解かれる。波動関数を原子局在基底の一次結合として表すと，次のような行列の固有値問題が得られる（Hartree-Fock-Roothaan 方程式，以下 HFR 方程式と略記）。

$$\hat{H}(\{R\},\{\alpha\},\{Q\},\{C\})C(\{R\},\{\alpha\},\{Q\}) = \hat{S}(\{R\},\{\alpha\},\{Q\})C(\{R\},\{\alpha\},\{Q\})e(\{R\},\{\alpha\},\{Q\}) \quad (2)$$

ここで，$\hat{H}$ はハミルトニアン行列，$\hat{S}$ は重なり積分行列，$C$ は波動関数（原子局在基底の一次結合係数），$e$ は固有値である。また，{R}は原子核の位置，{α}は基底関数の空間的な広がりを指定する係数(軌道指数)，{Q}は量子数であり，原子基底はこれらを用いて表現される。

HFR 方程式を解析式によって直接表現すると，これは，{R}，{α}，{Q}を変数にもつ多変数関数であり，超越関数を含む複雑な分子軌道方程式になる。これは一電子積分，二電子積分を構成する際の基底関数として，通常はガウシアン型(GTO)またはスレーター型軌道(STO)を用いるためである。解析式型基底は、数値計算上は効率的であるが、解析表現が複雑なため数学的な解析は困難を伴う。しかしながら，この方程式に含まれる変数について多項式近似を行い、簡易な表式を得ることが可能である(ワイエルシュトラスの近似定理)。この目的では，テイラー展開を行えば十分である。分子軌道方程式は原子座標，軌道指数，量子数を用いて表される高次代数方程式の連立系となる。その具体的な表現は数式処理を用いることで計算可能である[4-11]。なお分子軌道法における HFR 方程式の多項式近似という概念は，安井によって開発されたものである[6-9]。安井は，多項式近似した HFR 方程式を「分子軌道代数方程式」と名付けている。このことによって，分子積分の多項式近似の上に分子軌道の多変数理論を構築することができる。分子の性質は多変数関数として表現され，通常各変数間の連関を明確に追及することが可能になる。

通常の方法の電子状態計算は材料物性計算における順問題である。すなわち，物質の構造データの仮定値を入力し，電子状態計算を行い，全エネルギーが極小になるように構造を最適化し，最安定構造のもとで，種々の電子物性を求める。従来法には，次の課題がある。第一原理分子動力学法において，原子核は古典的なニュートン方程式に従うものであり，一方，波動関数は量子力学に従うと見なされる。その基礎となる理論が，いわゆる断熱近似である。断熱近似を適用することにより，波動関数と原子核とは，別々のモデルで取り扱われる。したがって，従来法では，緩和法による波動関数最適化と，原子配置の最適化という二つの計算フェーズを交互に繰り返しながら計算を進める。この方法は，数値的に安定である，とされる。しかし，波動関数と原子核配置の最適値を同時に求めるという見地からすると，回り道の多い，緩和問題の解法として非効率な手段になることがある。例えば，原子核と波動関数の自由度を分離して扱っているため，原子核と波動関数のダイナミクスが強く結合するような物理現象は，従来法では扱いにくいのである。これに対して、逆問題は，望ましい電子物性を与えるような物質構造を探求するということになる。従来法を用いて逆問題を検討する際に，まず物質の構造を仮定し，これに対する電子物性を求め，計算結果が望ましい値に近づくように仮定を少しずつ変化させながら繰り返し計算を行う。すなわち，仮定した構造を少しずつ変え，順問題を繰り返しとくことによって，

試行錯誤的に逆問題の解へいたる，という道筋しかもたない。従来法の電子状態計算において，固有値計算・自己無撞着場計算・原子核系の構造緩和計算の順という一定の順序にしたがって，複数の緩和計算が入れ子構造になっている。未知変数は，いかなる場合でも，内側から外側のループまで，決まった順番にしたがって決定しなければならない。それが，従来法における逆問題の解法が，順問題の試行錯誤的な繰り返しになる理由である。

## 2. 記号的-数値的第一原理分子動力学・分子軌道法

われわれは，この状況を鑑み，次の手法を提案する。
「ハートリー・フォック・ロータン方程式を連立多変数多項式系として近似表現し，さらに，この連立多項式系を数式処理によって『より解きやすい形』に変更し，その根を記号的-数値的に求めることによって固有状態を求める。」

本研究における課題は，分子軌道方程式を多項式近似した後に，どのように方程式の解を求め，有意義な情報を導くかということである。一般に，有限個数の解を持つ連立多項式系を解こうとする場合には，浮動小数点近似に基づき解は数値的に計算される。しかしながら，数値誤差等の影響によって，純粋に数値計算的な手法はしばしば不安定になり，その不安定性が何処に向かうのか予測が困難になることが多い。そこで，記号計算(数式処理)と数値計算を組み合わせた解析方法(Symbolic-numeric solving)が提唱されている。Symbolic-numeric solving では，記号計算によって，前処理として，記号計算を連立多項式系に適用し，同一の根を持ち，かつ安定かつ容易に数値計算が可能な連立方程式を作成する。さらに，この時点で，解の特徴，例えば，解が存在するか，どういう幾何学的構造を持つかなどを判断する。その後，数値計算に移行するのである。数学的背景は[12-15]を参照のこと。計算化学分野における記号計算の適用事例は，[16]にレビューがある。

本手法は，HFR 方程式を解くために、Symbolic-numeric solving の幾つかの手法を利用し、連立多変数多項式系として近似した HFR 方程式は，同一の根を持つ方程式系に変換される。一つの方法として，三角化のアルゴリズム(参考文献[12,13,14,15])を適用する。このアルゴリズムにおいて次のような変換が行われる。

出発点となる方程式系(The starting equations)$f_1, …, f_n$

$$f_1(x_1, x_2, …, x_n) = 0$$
$$f_2(x_1, x_2, …, x_n) = 0$$
$$\vdots$$
$$f_n(x_1, x_2, …, x_n) = 0$$

(3)

→ $f_1, …, f_n$ の辞書式順序グレブナー基底(Gröbner bases with lexicographic monomial order) $\{g_i\}$

$$g_1(x_1) = 0 \tag{4}$$
$$\vdots$$
$$g_{2\_1}(x_1, x_2) = 0$$
$$\vdots$$
$$g_{2\_m(2)}(x_1, x_2) = 0$$
$$g_{3\_1}(x_1, x_2, x_3) = 0$$
$$\vdots$$
$$g_{n\_1}(x_1, \ldots, x_n) = 0$$
$$\vdots$$
$$g_{n\_m(n)}(x_1, \ldots, x_n) = 0$$

→ 三角化された多項式系(Triangular sets of polynomials)  $\{t_i\}$
$$t_1(x_1) = 0 \tag{5}$$
$$t_2(x_1, x_2) = 0$$
$$\vdots$$
$$t_n(x_1, x_2, \ldots, x_n) = 0$$

文献[12,13]に示されたアルゴリズムでは，出発点の方程式系$f_1, \ldots, f_n$からまず辞書式順序のグレブナー基底$\{g_i\}$をつくる。このグレブナー基底も方程式系$f_1, \ldots, f_n$と同等の根を持ち，数値的に「より解きやすい」形式になっている。すなわち未知変数の少ない多項式からはじまり，未知変数の多い多項式にいたるように，多項式系が整理されている。しかし，多項式の数が出発点の方程式系よりも一般に大きくなる。この段階で，根の探索を実施することもできる。ここで，グレブナー基底に対し，三角化のアルゴリズムを適用すると，(5)に示すように，それぞれ未知変数$x_1$，未知変数$x_1, x_2$，…，未知変数$x_1, x_2, \ldots, x_n$をもつ n 個の多項式から成る三角多項式系$t_1, t_2, \ldots, t_n$を得る。なお，出発点の方程式系$f_1, \ldots, f_n$は，通常，複数の根を有する。そのため，方程式系$f_1, \ldots, f_n$の全零点を与えるために，三角多項式系$t_1, t_2, \ldots, t_n$が複数必要になることがあるが，参考文献の三角化アルゴリズムでは，必要な三角多項式系すべてを生成することが可能である。三角化された方程式系は，グレブナー基底系よりも，方程式のサイズが小さく，解きやすいものである。こうして得られた三角化方程式系$t_1, t_2, \ldots, t_n$を用い，未知数$x_1, x_2, \ldots, x_n$をひとつずつ順に求めていくことができる。すなわち，HFR 方程式を多項式系とみなし，数式処理によって三角化することで，未知数である固有値，波動関数，各種パラメータの値を，順を追って決定することができるようになる。この際，数値計算に必要なのは，一変数に対するニュートン法あるいはその類似の方法だけである。

さらに、この種の方程式を解くためにSymbolic-numeric solvingの別の方法も利用できる。この解法の基礎は，Stickelbergerの定理である。HFR 方程式を変数$X_1, X_2, \ldots, X_m$で表される多項式の集合$f_1, \ldots, f_m$と見る。この多項式の集合は，数学的な見地では，可換環 R = k$[X_1, \ldots, X_m]$上の零次元イデアル I を構成し，イデアル I の零点（イデアル I を連立方程式と見たときの根）は剰余環 A＝R/I に対応する。kは，多項式環の係数体であり，われわれの場合，有理数体もしくは実数体としてよい。A はk上有限次元のベクトル空間であり，このベクトル空間の基底は，$X_1, X_2, \ldots, X_m$の単項式によって表される。したがって，A=R/I

において，各基底に $X_1, X_2, ..., X_m$ を演算した結果も，A の基底の線形結合になる。すなわち，$X_1, X_2, ..., X_m$ の積演算は，基底間の線形変換行列 $m_h$（$h = X_1, X_2, ..., X_m$）として表されるのである。剰余環 A＝R/I の基底および変換行列 $m_h$ は，イデアル I のグレブナー基底を利用して求めることができる。Stickelberger の定理は，「行列 $m_h$ の固有ベクトル $v_\xi$ とイデアル I の零点 $\xi = (\xi_1, ..., \xi_m)$ が一対一の対応をする」ということを主張するものである。その対応は

$$m_{x_i} \cdot v_\xi = \xi_i \cdot v_\xi \qquad (6)$$

である。ここで，固有ベクトル $v_\xi$ は，すべての $m_{x_i}$ について共通になる。（文献 14，p101-130, "From Enumerative Geometry to Solving Systems of Polynomial Equations", Frank Sottile を参照のこと。）零点 $\xi = (\xi_1, ..., \xi_m)$ を求めるための数値計算は次のように行う。どれか一つの $X_i$ に対して，上の固有値方程式を数値計算し，$v_\xi$ を求める。そうして得られた $v_\xi$ を他の $m_{x_j}$ ($j \neq i$) に掛け，固有値 $\xi_j$ ($j \neq i$) を求めることによって，イデアル I の零点 $\xi = (\xi_1, ..., \xi_m)$ が計算される。

　ここで，エネルギー汎関数，波動関数の規格化条件，HFR 方程式は波動関数（LCAO の係数）と固有値については多項式である。ここで，これらの解析式は，分子積分を用いて構成されるが，分子積分は，そのパラメータについては一般に解析関数で表され，多項式とはならない。[文献１,3-7]。そこで，分子積分は，そのパラメータに関する近似多項式によって置き換えるものとする。そうすることによって，波動関数と固有値ばかりでなく分子積分のパラメータまで含めた連立多変数多項式系、すなわち分子軌道代数方程式を作ることができる。ハートリー・フォック・ロータン方程式に他の拘束条件が加わる場合も，拘束条件の近似多項式を作り，方程式系に追加すればよい。方程式系の係数（実数）は有理数化しておけば，記号的に方程式系を変形する際に任意精度計算を行うことによって精度の低下を避けることができる。

　この方法の一つの利点は次のようなものである。従来法における入力データは分子構造であり，出力は電子状態である。しかし，本法では，入力データは，分子構造に限定されない。HFR 方程式および付随の拘束条件に現れる任意の変数を選び入力値とすることができる。解こうとする問題が適切に設定されていれば，該入力値に対して，他の未知変数がどのような値をとるのかを計算できる。問題が適切に設定されているかどうか，すなわち連立多変数多項式系に解があるかどうかは，代数学のイデアル理論に基づき，連立多変数多項式系のグレブナー基底の零値を与える解集合が存在するか否かという条件によって判定する。

　本手法のフローチャートを順に解説する。
１．エネルギー汎関数の解析式を求める。
　　変数として，波動関数（LCAO の係数），分子積分のパラメータ（原子座標，原子基底の減衰指数），その他の物理量を用いる。
２．拘束条件を解析式化する。汎関数および拘束条件中の係数（実数）は有理数化しておけば，記号的に方程式系を変形する際に精度の低下を避けることができる。
３．エネルギー汎関数，拘束条件は，固有値および波動関数（LCAO の係数）については多項式であるが，分子積分に含まれるパラメータに関しては一般に多項式にならない。そこで，分子積分は，その中に含まれるパラメータに関する近似多項式に置き換える

ものとする。例えば，分子積分のパラメータに，ある基準点（基準数値）を設定し，その基準点の周囲でテイラー展開することで，近似多項式を作ることができる。

4．多項式近似されたエネルギー汎関数と拘束条件について，エネルギー汎関数の極小条件に基づき，各変数に関する微分の解析式をつくる。これが解くべき多項式方程式系である。
5．作業4で作られた多項式方程式系を数式処理し，同一の根を持つ連立多項式系に変換する。その際，多項式方程式系は，一旦グレブナー基底に変換し，ここで，解の存在可能性を検討する。解が孤立点の集合である場合，すなわち，解が数値として求められる場合は，グレブナー基底を三角化する。あるいは，グレブナー基底を利用し，Stickelbergerの定理に基づき根の探索を固有値問題の形態に置き換える。
6．作業5で得られた方程式系を，数値的に解くことによって，この方程式系の根を求める。そうすることで，電子状態およびその他の未知変数の値が求められる。

## 3. 計算例

計算例を示す。以下，単位は原子単位系（atomic units）とする。まず，自己無撞着電子状態計算が可能であることを見る。対象として，水素分子を扱う。以下の解説では，最も単純な分子である水素原子の計算を示すが，本方法の適用対象は，二原子系・二電子系に限定されるものではないことに注意する。水素分子を取り上げた理由は，この系は，単純ではあるが，物質中に働く電子－電子間，電子－核間，核－核間の量子論的な相互作用を全て含むものであって，一般の多原子・多電子系の雛型と見なすことができるからである。水素分子の計算に必要な分子積分(一中心積分，二中心積分)はSTO基底を用いて作成した。エネルギー汎関数は，水素A，B其々の一中心積分，二中心積分，また波動関数の係数a，b，c，dの解析式であり，核間距離Rについては指数関数を含む。二中心積分の一例を(7)に示す。これは1s軌道間の二電子反発積分であり，クーロン型と呼ばれるものである。

その定義は $\iint drdr' \frac{\phi^{1s}(r-R_A)\phi^{1s}(r-R_A)\phi^{1s}(r'-R_B)\phi^{1s}(r'-R_B)}{|r-r'|}$ であり、[1s(A)1s(A)||1s(B)1s(B)]と表すことができる。ここで1s(A),1s(B)は原子A,Bに中心を持つ1s軌道の意である。式中、za, zb, zc, zdは四つの1s軌道の軌道指数，Rは核間距離，$E^{R(za+zb)}$はexp((za + zb)R)の意である。二電子反発積分として、この他に、Exchange型[1s(A)1s(B)||1s(A)1s(B)]やhybrid型[1s(A)1s(B)||1s(A)1s(B)]型の積分もある。一般には，分子積分は，この式よりもさらに複雑な超越関数を含む式となる。STO基底を用いる場合は，指数関数や指数関数積分などが現れ，Exchange型反発積分に至っては無限級数和となる。しかし，如何なる複雑な解析式であっても，これを有限次数の多項式で近似することで，数式処理が容易になる。しかも、STO基底は，原子核近傍と遠距離の局在軌道の物理的な性質をGTO軌道に較べて精確に表現することが可能であり，原子座標に関するテイラー展開によって分子方程式を多項式化するという目的に対して有利である。この理由から、本研究ではSTO基底を用いた。ただし，以下に述べる処方はGTO基底に対しても、また、AM1，PM3，強結合モデル等の半経験的手法に対しても適用可能である。

$$[1s(A)1s(A)|1s(B)1s(B)]= \tag{7}$$

$$\frac{64\,z_a^{3/2}\,z_b^{3/2}\,z_c^{3/2}\,z_d^{3/2}}{E^R\,(z_a+z_b)^3\,(z_c+z_d)^3} - \frac{32\,z_a^{3/2}\,z_b^{3/2}\,(z_a+z_b)\,z_c^{3/2}\,z_d^{3/2}}{E^R\,(z_c+z_d)^2\,(z_a+z_b-z_c-z_d)^2\,(z_c+z_d)^2\,(z_a+z_b+z_c+z_d)}$$

$$- \frac{32\,z_a^{3/2}\,z_b^{3/2}\,z_c^{3/2}\,z_d^{3/2}\,(z_c+z_d)}{E^R\,(z_a+z_b)^2\,(z_a+z_b)^2\,(z_a+z_b-z_c-z_d)^2\,(z_a+z_b+z_c+z_d)^2}$$

$$- \frac{64\,z_a^{3/2}\,z_b^{3/2}\,z_c^{3/2}\,z_d^{3/2}\,(z_c+z_d)\,(3\,z_a^2+6\,z_a z_b+3\,z_b^2-z_c^2-2\,z_c z_d-z_d^2)}{E^R\,(z_a+z_b)^3\,R\,(z_a+z_b-z_c-z_d)^3\,(z_a+z_b+z_c+z_d)^3}$$

$$+ \frac{64\,z_a^{3/2}\,z_b^{3/2}\,(z_a+z_b)\,z_c^{3/2}\,z_d^{3/2}\,(-z_a^2-2\,z_a z_b-z_b^2+3\,z_c^2+6\,z_c z_d+3\,z_d^2)}{E^R\,(z_c+z_d)^3\,R\,(z_a+z_b-z_c-z_d)^3\,(z_c+z_d)^3\,(z_a+z_b+z_c+z_d)^3}$$

　　本方法の順問題的第一原理分子電子動力学法に対する適用事例として，構造最適化（原子間距離の最適化）と UHF 電子状態計算を同時に実行する。

　ここで，水素 A, B の位置を$R_A$ および$R_B$とする。核間距離を$r = R_A - R_B$ とする。また，$x_A = |x - R_A|$，$x_B = |x - R_B|$という記法を用いる。UHF 計算を行うので，上向き・下向きのスピンに対する試行波動関数を

$$\phi_{up}(x) = a\exp(-x_A) + b\exp(-x_B) \quad (8)$$

および

$$\phi_{down}(x) = c\exp(-x_A) + d\exp(-x_B) \quad (9)$$

とする。それぞれに対応する固有値を ev，ew とする。なお，基底（軌道指数 1）を用いた計算値は実験値と良く一致するものではない。実測を再現するためには，軌道指数を適切な値に設定する必要がある。軌道指数 1 の基底を用いるのは，単に，数式処理のコストを削減する目的である。

　エネルギー汎関数は，まず解析式を数式処理によって作成したのち，$R_0=7/5$ の位置で，核間距離 r に関し四次のテイラー展開を行い，多項式化した。多項式化した波動関数の実数係数は，絶対値 1/1000 以下のものを切り捨て，有理数で近似した。これを(10)に示す。(10)中の記号は次の通りである。(a,b)は上向きスピン電子の LCAO 係数，(c,d)は下向きスピン電子の LCAO 係数，ev は上向きスピン電子のエネルギー固有値，ew は下向きスピン電子のエネルギー固有値，r は核間距離である。

Ω＝(3571 - 1580*a^2 - 3075*a*b - 1580*b^2 - 1580*c^2 + 625*a^2*c^2 +　　(10)
1243*a*b*c^2 + 620*b^2*c^2 - 3075*c*d + 1243*a^2*c*d + 2506*a*b*c*d
+ 1243*b^2*c*d - 1580*d^2 + 620*a^2*d^2 + 1243*a*b*d^2 +
625*b^2*d^2 + 1000*ev - 1000*a^2*ev - 1986*a*b*ev - 1000*b^2*ev +
1000*ew - 1000*c^2*ew - 1986*c*d*ew - 1000*d^2*ew - 5102*r +
332*a^2*r + 284*a*b*r + 332*b^2*r + 332*c^2*r + 43*a*b*c^2*r +
20*b^2*c^2*r + 284*c*d*r + 43*a^2*c*d*r + 80*a*b*c*d*r +
43*b^2*c*d*r + 332*d^2*r + 20*a^2*d^2*r + 43*a*b*d^2*r -
63*a*b*ev*r - 63*c*d*ew*r + 3644*r^2 + 75*a^2*r^2 + 724*a*b*r^2 +
75*b^2*r^2 + 75*c^2*r^2 - 401*a*b*c^2*r^2 - 124*b^2*c^2*r^2 +
724*c*d*r^2 - 401*a^2*c*d*r^2 - 1372*a*b*c*d*r^2 - 401*b^2*c*d*r^2 +

75*d^2*r^2 - 124*a^2*d^2*r^2 - 401*a*b*d^2*r^2 + 458*a*b*ev*r^2 + 458*c*d*ew*r^2 - 1301*r^3 - 69*a^2*r^3 - 303*a*b*r^3 - 69*b^2*r^3 - 69*c^2*r^3 + 146*a*b*c^2*r^3 + 42*b^2*c^2*r^3 - 303*c*d*r^3 + 146*a^2*c*d*r^3 + 618*a*b*c*d*r^3 + 146*b^2*c*d*r^3 - 69*d^2*r^3 + 42*a^2*d^2*r^3 + 146*a*b*d^2*r^3 - 139*a*b*ev*r^3 - 139*c*d*ew*r^3 + 185*r^4 + 12*a^2*r^4 + 39*a*b*r^4 + 12*b^2*r^4 + 12*c^2*r^4 - 17*a*b*c^2*r^4 - 4*b^2*c^2*r^4 + 39*c*d*r^4 - 17*a^2*c*d*r^4 - 86*a*b*c*d*r^4 - 17*b^2*c*d*r^4 + 12*d^2*r^4 - 4*a^2*d^2*r^4 - 17*a*b*d^2*r^4 + 13*a*b*ev*r^4 + 13*c*d*ew*r^4)/1000

この近似の精度についてまず議論する。多項式近似適用前および後の汎関数に，値 a=b=c=d=1，ev=ew=0 を代入し，核間距離 r についてプロットしたものを図 2 に示す。両者が概ねよい一致を示すのは r が 1~2 atomic unit の範囲であって，r がこの有効範囲を超えた場合，多項式近似は不適切であり，別の展開中心点 $R_0$ を使用する必要がある。テイラー展開の次数を上げることによっても，近似が有効な r の範囲を広げることができる。

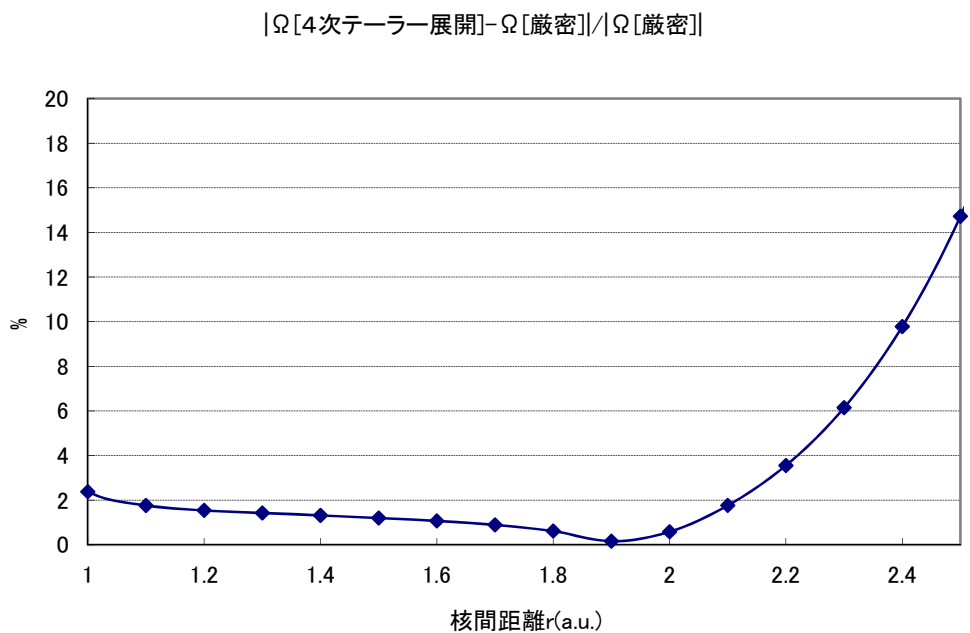

図 2　水素分子のエネルギー汎関数のテイラー展開と厳密式とのずれ。

ここで，水素分子の対称性を利用し，波動関数を対称・反対称波動関数の線形結合

$$\phi_{up}(x) = \text{t}(\exp(-x_A) + \exp(-x_B)) + \text{s}(\exp(-x_A) - \exp(-x_B)) \tag{11}$$

$$\phi_{down}(x) = \text{u}(\exp(-x_A) + \exp(-x_B)) + \text{v}(\exp(-x_A) - \exp(-x_B)) \tag{12}$$

とする。これは，

$$a = \text{t} + \text{s}, b = \text{t} - \text{s}, c = \text{u} + \text{v}, d = \text{u} - \text{v} \tag{13}$$

という変換である。

そうすると，HFR 方程式は(14)に示す連立多項式方程式になる。 (t,s) は上向きスピン

電子のLCAO係数，(u,v)は下向きスピン電子のLCAO係数，evは上向きスピン電子のエネルギー固有値，ewは下向きスピン電子のエネルギー固有値，rは核間距離である。

① $\dfrac{\partial \Omega}{\partial a} + \dfrac{\partial \Omega}{\partial b} = 0 \rightarrow$ (14)

32*s*u*v*r^4-336*s*u*v*r^3+992*s*u*v*r^2-160*s*u*v*r+40*s*u*v-324*t*u^2*r^4+2572*t*u^2*r^3-6448*t*u^2*r^2+584*t*u^2*r+19936*t*u^2+156*t*v^2*r^4-1068*t*v^2*r^3+2248*t*v^2*r^2-80*t*v^2*r-32*t*v^2+26*t*ev*r^4-278*t*ev*r^3+916*t*ev*r^2-126*t*ev*r-7972*t*ev+126*t*r^4-882*t*r^3+1748*t*r^2+1896*t*r-12470*t=0

② $\dfrac{\partial \Omega}{\partial a} - \dfrac{\partial \Omega}{\partial b} = 0$

→156*s*u^2*r^4-1068*s*u^2*r^3+2248*s*u^2*r^2-80*s*u^2*r-32*s*u^2-52*s*v^2*r^4+236*s*v^2*r^3-32*s*v^2*r^2-104*s*v^2*r+48*s*v^2-26*s*ev*r^4+278*s*ev*r^3-916*s*ev*r^2+126*s*ev*r-28*s*ev-30*s*r^4+330*s*r^3-1148*s*r^2+760*s*r-170*s+32*t*u*v*r^4-336*t*u*v*r^3+992*t*u*v*r^2-160*t*u*v*r+40*t*u*v=0

③ $\dfrac{\partial \Omega}{\partial c} + \dfrac{\partial \Omega}{\partial d} = 0$

→156*s^2*u*r^4-1068*s^2*u*r^3+2248*s^2*u*r^2-80*s^2*u*r-32*s^2*u+32**t*v*r^4-336*s*t*v*r^3+992*s*t*v*r^2-160*s*t*v*r+40*s*t*v-324*t^2*u*r^4+2572*t^2*u*r^3-6448*t^2*u*r^2+584*t^2*u*r+19936*t^2*u+26*u*ew*r^4-278*u*ew*r^3+916*u*ew*r^2-126*u*ew*r-7972*u*ew+126*u*r^4-882*u*r^3+1748*u*r^2+1896*u*r-12470*u=0

④ $\dfrac{\partial \Omega}{\partial c} - \dfrac{\partial \Omega}{\partial d} = 0$

→-52*s^2*v*r^4+236*s^2*v*r^3-32*s^2*v*r^2-104*s^2*v*r+48*s^2*v+32*s*t*u*r^4-336*s*t*u*r^3+992*s*t*u*r^2-160*s*t*u*r+40*s*t*u+156*t^2*v*r^4-1068*t^2*v*r^3+2248*t^2*v*r^2-80*t^2*v*r-32*t^2*v-26*v*ew*r^4+278*v*ew*r^3-916*v*ew*r^2+126*v*ew*r-28*v*ew-30*v*r^4+330*v*r^3-1148*v*r^2+760*v*r-170*v=0

⑤ $\dfrac{\partial \Omega}{\partial (ev)} = \langle \phi_{up} | \phi_{up} \rangle - 1 = 0$

→-13*s^2*r^4+139*s^2*r^3-458*s^2*r^2+63*s^2*r-14*s^2+13*t^2*r^4-139*t^2*r^3+458*t^2*r^2-63*t^2*r-3986*t^2+1000=0

⑥ $\dfrac{\partial \Omega}{\partial (ew)} = \langle \phi_{down} | \phi_{down} \rangle - 1 = 0$

→13*u^2*r^4-139*u^2*r^3+458*u^2*r^2-63*u^2*r-3986*u^2-13*v^2*r^4+139*v^2*r^3-458*v^2*r^2+63*v^2*r-14*v^2+1000=0

⑦ $\frac{\partial \Omega}{\partial r} = 0$

→312*s^2*u^2*r^3-1602*s^2*u^2*r^2+2248*s^2*u^2*r-40*s^2*u^2-104*s^2*v^2*r^3+354*s^2*v^2*r^2-32*s^2*v^2*r-52*s^2*v^2-52*s^2*ev*r^3+417*s^2*ev*r^2-916*s^2*ev*r+63*s^2*ev-60*s^2*r^3+495*s^2*r^2-1148*s^2*r+380*s^2+128*s*t*u*v*r^3-1008*s*t*u*v*r^2+1984*s*t*u*v*r-160*s*t*u*v-648*t^2*u^2*r^3+3858*t^2*u^2*r^2-6448*t^2*u^2*r+292*t^2*u^2+312*t^2*v^2*r^3-1602*t^2*v^2*r^2+2248*t^2*v^2*r-40*t^2*v^2+52*t^2*ev*r^3-417*t^2*ev*r^2+916*t^2*ev*r-63*t^2*ev+252*t^2*r^3-1323*t^2*r^2+1748*t^2*r+948*t^2+52*u^2*ew*r^3-417*u^2*ew*r^2+916*u^2*ew*r-63*u^2*ew+252*u^2*r^3-1323*u^2*r^2+1748*u^2*r+948*u^2-52*v^2*ew*r^3+417*v^2*ew*r^2-916*v^2*ew*r+63*v^2*ew-60*v^2*r^3+495*v^2*r^2-1148*v^2*r+380*v^2+740*r^3-3903*r^2+7288*r-5102=0

まず，本法で実際に第一原理計算が可能なことを示すため、R=7/5 を仮定して解く。方程式(14)において，

$$\frac{\partial \Omega}{\partial r} = 0 \tag{15}$$

という式を

$$5r - 7 = 0 \tag{16}$$

と置き換える。この方程式系から生成された辞書式順序グレブナー基底は(17)である。単項式順序は，s<t<u<v<ev<ew<r の順とする。今後の計算ではこの単項式順序を使う。そのため，グレブナー基底(17)の各要素において，変数は，これとは逆に，r, ew, ev, v, u, t, s の順序で現れる。HFR 方程式の段階では曖昧であった変数間の相互関連が，順次抽出されてくることに注意しよう。

J[1]=r-1.4　　　　　　　　　　　　　　　　　　　　　　　　　　　　(17)
J[2]=0.00000000000000004492367995028017983415375254583712005086 9800531719*ew^6+0.0000000000000000837848476849489179918230616638 7981324102296825 5965*ew^5+0.00000000000000005522431406589942780 2789765894852689587104815167352*ew^4+0.00000000000000014691161 15773407851793157859874999641802949934 6849*ew^3+0.0000000000000 0000119793511852426815972991408741146613467174047 26834*ew^2-0.00 0000000000000000839869963562374452466573406 5856989 58493547765 49449*ew-0.00000000000000000000372304090531559502639835221836 91557632339096670694
J[3]=0.00000000013962876535705715746622071381953077198921180700 71*ev+0.0000003217365693623504793206423800429601988107513122 5526*ew^5+0.000000332916524777360052071820343090980816 28696594819331*ew^4+0.0000000658911032364830212609562005105 55934469425614980

251*ew^3-0.00000001450219479632473049589549258544954048787083075
49*ew^2-0.00000002589663947030314527682380814926503763029302261
6944*ew+0.000000000508191975132258812443532309139696862135914949
4118136

J[4]=0.0035337421249281116351414330877752843125543561935616*v*ew^4+0.0021801710635532512041192523955501421176113052868825*v*ew^3+0.000246834120767041208618495460378545445015627301666012*v*ew^2-0.0000145546516842582525115018507036947117736796217302905*v*ew-0.00000075203364912461984107689153015151456273923670860722*v

J[5]=0.0000000000000246918226391259732719708078543665008858005230838289*v^2-0.000000000047190217319792339322638580120931177672487391351614*ew^5-0.0000000000742337258614370125563724850891349127581737428597544*ew^4-0.000000000037044087563918384281802854504031102187085751576865*ew^3-0.0000000000055056937948011551503818116670012263028105621886299*ew^2+0.00000000000028924710974820572210152554673921687498485750622516*ew-0.000000000000485199500275950858789604761685291061709957821845754

J[6]=0.00082444529533914004801217492520348576142750871636346*u*ew^4+0.00154024712861099035299602345961334386566856243445*u*ew^3+0.001018615510141267404109756213974524296027592026266*u*ew^2+0.0002733014715179344910880975554697358492670478484359*u*ew+0.0000231526498118394429461881625790391710631015256860316*u

J[7]=0.1550065208537365727827215504884042526908166686142*u*v*ew^2+0.0961244867459237984530250863300832527616301366651358*u*v*ew+0.0111781199009703988246113307469548776395281830706833*u*v

J[8]=0.00000000000002469182263912597327197080785436650088580523 0838289*u^2+0.0000000000006618541962195662625184586935775373890054531569619*ew^5+0.000000000001041145935596238702817938098996908688033630169190868*ew^4+0.00000000000051955227570060962098498758121689908883959384515193*ew^3+0.0000000000000772186853155438576293540302694999396484080162796235*ew^2-0.0000000000000405760577886977743474154857219642292365593374753795*ew-0.0000000000000000023368741490554450106147082788613043426219204168902

J[9]=0.0619709181331632165671838700351535038880015135145687*t*ew^4+0.077869672985153610660198919208791058623344646828889*t*ew^3+0.02952928754574380870962044724506192695086361951332133*t*ew^2+0.003217837502146734951260237098620435482764135949341134*t*ew+0.00004345685303182307000128985173460828330264279573540644*t

J[10]=0.05187413667297784999868866991024835274577850308566*t*v*ew^3+0.0329814165688916253656643121212067736852468257601442*t*v*ew^2+0.0042447701517134049943862889216523608128280851170331*t*v*ew+0.00005860046167413441588422093257956926678479916969229955*t*v

J[11]=0.0217917033542653598802626469372919880768213136 85722*t*u*ew^3+0.027041037783077439842866167839432112120 82455989442*t*u*ew^2+0.009960198861693293199753059700357 0949767046400427524*t*u*ew+0.00097550648450924773054739454 1626918868204022901 90827*t*u

J[12]=0.000000000000000246918226391259732719708785436650 0885805230838289*t^2-0.000000000001989710106051965210739512 5478208499820923810270269*ew^5-0.000000000001702557008975665 5354942878105283795981119659773602*ew^4-0.0000000000002348359 2200304199348742150925412718812994059865964*ew^3+0.00000000000 019673727812808485015748182293664380763085511692487*ew^2+0.00 000000000001978607845702383305259202618110582102686639053576 6*ew-0.0000000000000044219079290739404492269360637167466142801 279295227

J[13]=0.056084833175126678686494811325106385425518369 783604*s*ew^2+0.034127439269225675895016122448171947840 127214030889*s*ew-0.000662823004416982327050787731273824 869454021011201 19*s-0.0408938850202494659735531919384785903099600793996 38*t*u*v*ew-0.029010301971762382682320743416650072000710 482565746*t*u*v

J[14]=0.0217917033542653598802626469372919880768213136 85722*s*v*ew-0.000410528622714882375331121888578012581137 75891824163*s*v-t*u*ew^2-0.915232433317809793085161470678 69265402585316644918*t*u*ew-0.1827983573960928404931571886 04362641091180814 04686*t*u

J[15]=0.082689621743132526006561521010847989515268982 874621*s*u*ew+0.051874136672977849998688669910248352745 77850308566*s*u+t*v*ew^2+0.450881584834248624146025087680 1902904814470 8971225*t*v*ew+0.006817682348103806117952268152901245007 4191772534997*t*v

J[16]=0.061970918133163216567183870035153503888001513 514567*s*u*v+7.54843402550665153673201812730961379367248 454713989*t*ew^3+9.297759662579435885826123392802703170 3300204455612*t*ew^2+2.933336922081654496752241011010907 46851954196539922*t*ew+0.043698233453379147442339681621264 612101282386252875*t

J[17]=0.155006520853736572782721550488404252690816668 61422*s*t-u*v*ew-0.310065945021198217903859863692486629 8895502223605*u*v

J[18]=0.00000000000000024691822639125973271970878543665 00885805230838289*s^2+0.0000000000014186637003179206604978 647501193712331871881329982*ew^5+0.00000000001213923485139 370932693489804708248 0718477074921104*ew^4+0.000000000001674380589730518276554 0212510939603762998421893102*ew^3-0.000000000000140273718 3325013576178 8614236060225048185772213358*ew^2-0.000000000000141074778 648629063444287891909984518147600123 08984*ew-0.00000000000001865792945 3315597187256856909000195939743454954377

グレブナー基底(17)を三角化すると方程式(18)となる。ここには五つの三角化された連立方

程式が含まれる．一つの連立方程式は，七つの式を含み，変数は，r, ew, ev, v, u, t, s の順序で追加されていく．

[1]:                                                                                                (18)
  _[1]=r-1.4
  _[2]=0.082689621743132526006561521010847989515268982874621*ew
     +0.051874136672977849998688669910248352745778503085666
  _[3]=0.00000000013962876535705715746622071381953097719892118070071*ev
     +0.0000000000021872894505215127104699976329395744001395648217513
  _[4]=v
  _[5]=0.000000000000002469182263912597327197080785436650088580523083289*u^2
     -0.00000000000000070387276012025253831506975421079541060390364383305
  _[6]=t
  _[7]=0.000000000000002469182263912597327197080785436650088580523083289*s^2
     -0.0000000000000050186141759442129979160381879133091558873472043949
[2]:
  _[1]=r-1.4
  _[2]=ew-0.018838757853893431115706795065601087293579573640499
  _[3]=ev-0.018838757853893431115706795065601087293579573640499
  _[4]=v^2-2.032500495930121006949525695684269746961818258683366
  _[5]=u
  _[6]=t
  _[7]=0.000000000000002469182263912597327197080785436650088580523083289*s^2
     -0.0000000000000050186141759442129979160381879133091558873472043949

[3]:
  _[1]=r-1.4
  _[2]=ew^2+0.6201318900423964358077197273849732597791004444721*ew
     +0.072113868754708844917382916961852739039135803532834
  _[3]=ev-ew
  _[4]=v^2-14.809950234281971339519059215077293150891773692539334*ew
     -5.777653339333545581487020716322560290468549512904111
  _[5]=u^2+2.077131292262844160250221216181327899549542410218544*ew
     +0.525266732317117545403256347138142609545167168240854
  _[6]=t^2+2.077131292262844160250221216181327899549542410218544*ew
     +0.525266732317117545403256347138142609545167168240854
  _[7]=s+8.556707658075617499772076904610507582973719342317464*t*u*v*ew
     +6.2483498532831756195102551269084011635816028155391949*t*u*v
[4]:
  _[1]=r-1.4
  _[2]=ew+0.6207549439835869009416791421320771179233231538022444

_[3]=ev+0.62075494398358690094167914213207711792332315380224
        _[4]=v
        _[5]=u^2-0.28506310384917288370291954789447830487339321330884
        _[6]=t^2-0.28506310384917288370291954789447830487339321330884
        _[7]=s
  [5]:
        _[1]=r-1.4
        _[2]=ew+0.01566503467196175486397478472081382667860308248l373
        _[3]=ev+0.62733551780076996060l0l56274085545071738486487756
        _[4]=v^2-2.03250049593012100694952569568426974696181825868366
        _[5]=u
        _[6]=t^2-0.28506310384917288370291954789447830487339321330884
        _[7]=s

このときの解は，複素数解も含むが，物理的に意味のある実数解のみを表 1 に示す。上向き・下向きスピンを持つ二つの電子が対称・反対称波動関数の二準位のどちらかに位置する，四つの組合せに対する波動関数が得られた。すなわち，基底状態と励起状態の両方が得られたことになる。

|  | 解1<br>電子1～対称軌道<br>電子2～対称軌道<br>基底状態 | 解2<br>電子1～反対称軌道<br>電子2～対称軌道<br>励起状態 | 解3<br>電子1～対称軌道<br>電子2～反対称軌道<br>励起状態 | 解4<br>電子1～反対称軌道<br>電子2～反対称軌道<br>励起状態 |
|---|---|---|---|---|
| s（電子1係数） | 0.00000 | −1.42566 | 0.00000 | −1.42566 |
| t（電子1係数） | −0.53391 | 0.00000 | −0.53391 | 0.00000 |
| u（電子2係数） | −0.53391 | −0.53391 | 0.00000 | 0.00000 |
| v（電子2係数） | 0.00000 | 0.00000 | −1.42566 | −1.42566 |
| ev（電子1固有値） | −0.62075 | −0.01567 | −0.62734 | 0.01884 |
| ew（電子2固有値） | −0.62075 | −0.62734 | −0.01567 | 0.01884 |
| r（核間距離） | 1.40000 | 1.40000 | 1.40000 | 1.40000 |
| Etot（全エネルギー） | −1.09624 | −0.49115 | −0.49115 | 0.15503 |

表1　方程式系(18)の実数解。電子1（上向スピン），電子2（下向スピン）である。

　基底状態のみを得るために，方程式(13)に，式
$$s = v = 0 \tag{19}$$
を付け加える。（これは，この事例のみに適用可能な手段である。一般に，基底状態は，全固有値の和が最小になるという条件を満たす解として得ることができる。基底状態を特定するためには，固有値のみをまず計算すればよい。このとき，辞書式順序のグレブナー基底またはこれを三角化した方程式系を作る際に，固有値のみを含む方程式群をまず作成する。そして，固有値のみの方程式を解けばよい。）こうすることで方程式系を簡単なものに置き換え解くことができる。このとき，核間距離 r 及び波動関数は同時に最適化される(カー・パリネロ型解法)。r に関する方程式を(20)，実数解を表2に示す。

  [1]:  (20)
8942144364*r^23-435341589039*r^22+9813157241157*r^21-134458128500631*r
^20+1251986164962728*r^19-8584760758387395*r^18+48176522279858253*r^1

7-254992901607817871*r^16+1360184656773665254*r^15-6685412705413184235*r^14+26848712421674517351*r^13-82265960807423324641*r^12+185370480318135661708*r^11-295651827763150999108*r^10+307426892321213994312*r^9-148683667595876075980*r^8-97338526988608612178*r^7+245772518836579791529*r^6-200002425723099153061*r^5+472986381792776357737*r^4+46006348188804187952*r^3-41646082527529600720*r^2+131189224005435784966*r-1869747053688110592=0

[2]:
10313892*r^15-376866027*r^14+6245669754*r^13-61144647973*r^12+387764699571*r^11-1646957525797*r^10+4691411679124*r^9-8760215434992*r^8+10281598671237*r^7-7316755042677*r^6+3784010771997*r^5-2194016637700*r^4-299532295668*r^3+1482785614608*r^2-746000940352*r+36100845312=0

[3]:
10313892*r^15-376866027*r^14+6245669754*r^13-61144647973*r^12+378596795571*r^11-1405917509797*r^10+1917133055124*r^9+9044372533008*r^8-55118065080763*r^7+103464030245323*r^6+92432281739997*r^5-770797010005700*r^4+1063674493728332*r^3+652030557238608*r^2-2854269358708352*r+1954998898509312=0

|  | r | ev | t |
|---|---|---|---|
| 実数解1 | -1.812 | -6.6 | 0.846 |
| 実数解2 | 1.652 | -0.578 | 0.545 |
| 実数解3 | 6.010 | -17.585 | 0.983 |

表 2　方程式系(20)の実数解

解は，正負の実数値および複素数値を含む。意味のある解（r＞0）は，二つあるが，テイラー展開の有効範囲内にある解は，r～1.6 のみである。この解と実測値(r～1.4)のずれは、第一に四次テイラー展開と数値係数の有理化の粗さに由来する数値誤差によるものである、また、軌道指数を最適化していない試行関数をもちいていることにもよる。

次に本法を用いた逆問題の解法例を示す。水素分子の占有，非占有状態の固有エネルギー差がある特定の値を持つ状況を想定し，その値の出るような核間距離を求めることを考える。この実施例は，例えば，格子定数を変化させて望ましいバンドギャップを求めるような「逆問題」の雛型である。今回は，RHF 計算を行う。占有・非占有状態の波動関数を

$$\phi_{occ}(x) = s(\exp(-x_A) + \exp(-x_B)) \quad (21)$$

$$\phi_{unocc}(x) = t(\exp(-x_A) - \exp(-x_B)) \quad (22)$$

とし，固有値を $e_{occ}$, $e_{unocc}$ とした。この場合，占有状態を決める方程式は，上にあげた UHF

順問題計算と同様の方式で得ることができる。さらに非占有状態を決める方程式と占有・非占有状態間の直交条件が追加される。このとき必要な方程式系は(23)に示す。式詳細は省略する。

$$\frac{\partial \Omega}{\partial s} = 0 \tag{23}$$

$$\frac{\partial \Omega}{\partial t} = 0$$

$$\frac{\partial \Omega}{\partial (e_{occ})} = \langle \phi_{occ} | \phi_{occ} \rangle - 1 = 0$$

$$\frac{\partial \Omega}{\partial (e_{unocc})} = \langle \phi_{unocc} | \phi_{unocc} \rangle - 1 = 0$$

$$\langle \phi_{occ} | \phi_{unocc} \rangle = 0$$

$$e_{occ} - e_{unocc} = E_{gap}$$

例として，$E_{gap}=e_{unocc}-e_{occ}=0.9$ を与える r を求めてみよう。実数解は，r＝－1.103, 0.307, 1.643, 3.958 である。テイラー展開の有効範囲にある正の実数解は r=1.643 のみである。r を連続的に変化させた場合の占有・非占有準位の固有値を図 3 に示すが，この図と比べて適正な結果が得られたことがわかる。

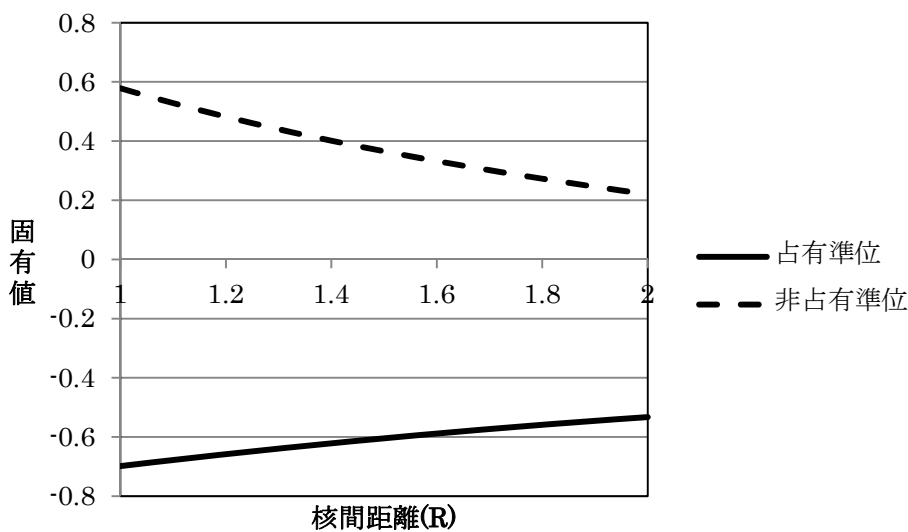

図3 水素分子の占有・非占有状態の固有値の核間距離依存性

では，この場合，構造は安定であるか。無論，力場計算を行えば，簡単に判定ができる。一方，Symbolic-Numeric 計算の立場からは，次のような判定を行うことができる。図 11 の方程式系に，条件 $E_{gap}=ew-ev=0.9$ 及び汎関数の核間距離 r に関する極値条件

$$\frac{\partial \Omega}{\partial r} = 0 \tag{24}$$

を追加する。このときのグレブナー基底の計算結果は{1}となる。グレブナー基底の零点は，

元来の多項式系の根を与えるが，そもそも多項式とみたとき{1}は零にならない。したがって，今回考える問題は解を持たないと結論できる

次に，これとは別種の数値解法である，Stickelberger の定理を利用した解法の実施例を示す。例題は，水素分子の基底状態の構造最適化であり，RHF 計算を実行する。試行関数を

$$\phi_{occ}(r) = t(\exp(-r_A) + \exp(-r_B)) \tag{25}$$

とする。固有値を ev，核間距離を r とする。前述の実施例と同様に，ハートリー・フォック方程式は，変数 t，ev，r で表される多項式となる。この多項式の集合は，数学的には多項式環 A = R[t,ev,r]における零次元イデアル I を構成し，イデアル I の零点は剰余環 A/I に対応する。R は実数体 R である。A は R 上有限次元のベクトル空間であり，このベクトル空間の基底は，t，ev，r の単項式によって表わされる。これを(26)に示す。

$$
\begin{array}{lllll}
b[1]=t*ev*r^3 & b[2]=t*r^4 & b[3]=t^3*ev & b[4]=t*ev^3 & b[5]=t^3*r \\
b[6]=t*ev^2*r & b[7]=t*ev*r^2 & b[8]=t*r^3 & b[9]=ev*r^3 & b[10]=r^4 \\
b[11]=t^3 & b[12]=t^2*ev & b[13]=t*ev^2 & b[14]=ev^3 & b[15]=t^2*r \\
b[16]=t*ev*r & b[17]=ev^2*r & b[18]=t*r^2 & b[19]=ev*r^2 & b[20]=r^3 \\
b[21]=t^2 & b[22]=t*ev & b[23]=ev^2 & b[24]=t*r & b[25]=ev*r \\
b[26]=r^2 & b[27]=t & b[28]=ev & b[29]=r & b[30]=1 \\
\end{array}
\tag{26}
$$

ベクトル空間 A/I の基底に対する変数 t の積演算を表す変換行列 $m_t$ を(27)に示す。

$$
\begin{pmatrix}
0. & 0. & 0. & 0. & 0. & 0. & 0. & 0. & 1. & 0. & 0. & 0. & 0. & 0. & 0. \\
0. & 0. & 0. & 0. & 0. & 0. & 0. & 0. & 0. & 1. & 0. & 0. & 0. & 0. & 0. \\
0. & 0. & 0. & 0. & 0. & 0. & 0. & 0. & 0. & 0. & 1. & 0. & 0. & 0. & 0. \\
0. & 0. & 0. & 0. & 0. & 0. & 0. & 0. & 0. & 0. & 0. & 1. & 0. & 0. & 0. \\
0. & 0. & 0. & 0. & 0. & 0. & 0. & 0. & 0. & 0. & 0. & 0. & 1. & 0. & 0. \\
0. & 0. & 0. & 0. & 0. & 0. & 0. & 0. & 0. & 0. & 0. & 0. & 0. & 0. & 1. \\
2.46857 & 0.250358 & 0.000433932 & -0.519297 & 0.000905017 & -0.229989 & 0.400711 & 0.0592776 & 0. & 0.00037686 & 0. & -0.0488928 & 0. & 0.0253457 & 0.0106224 \\
2.85865 & 1.8302 & 0.010766 & -3.25489 & 0.00562066 & 0.195796 & 0.755874 & 0.513375 & 0. & -0.0000321824 & 0. & 0.188933 & 0. & 0.177659 & 0.101971 \\
-3.28726 & 16.9846 & 0.815632 & -68.3604 & 0.199106 & -6.99935 & 17.2255 & 12.5718 & 0. & 0.010156 & 0. & 2.94616 & 0. & 8.19509 & 3.31843 \\
1.10471 & 0.0541875 & 0.0005962 & -0.43266 & 0.000241163 & 0.0161693 & 0.169585 & 0.0191714 & 0. & 4.36536 \times 10^{-6} & 0. & 0.00658837 & 0. & 0.0121792 & 0.00422577 \\
-41.1091 & 92.4474 & 0.457252 & -248.171 & 0.949199 & 7.93844 & -9.06291 & 33.1339 & 0. & 0.0550604 & 0. & 7.29159 & 0. & -1.91674 & 7.47555 \\
0.381952 & -0.482168 & -0.00366255 & 1.05379 & -0.00214058 & -0.266467 & 0.0658373 & -0.170307 & 0. & -0.0000384651 & 0. & -0.0484318 & 0. & 0.00667468 & -0.0375156 \\
-13.5886 & -1.50329 & 0.000693225 & 2.88068 & -0.00965266 & 1.64897 & -2.2783 & -0.319754 & 0. & -0.00357141 & 0. & 0.303898 & 0. & -0.125453 & -0.05279 \\
-9.85681 & -9.15049 & -0.0964526 & 22.4732 & -0.0359414 & -1.57568 & -3.20362 & -3.05264 & 0. & 0.00102063 & 0. & -1.41532 & 0. & -1.22295 & -0.657454 \\
480.008 & -277.275 & -3.39938 & 331.95 & 0.518864 & -172.803 & 89.1965 & -51.8746 & 0. & 0.670983 & 0. & -36.5975 & 0. & 14.0368 & 1.13365 \\
7.34299 & 6.6392 & 0.00938662 & -17.0623 & 0.0147453 & 1.43466 & 1.18689 & 1.56243 & 0. & -0.000518309 & 0. & 0.363976 & 0. & 0.111898 & 0.278794 \\
15.2609 & -1.66537 & -0.0325943 & 4.78209 & -0.0255023 & -8.07839 & 2.60648 & -0.91568 & 0. & 0.00589181 & 0. & -0.519365 & 0. & 0.0343565 & -0.229064 \\
4.37903 & 18.2502 & 0.31662 & -48.1151 & 0.0770211 & 4.75029 & 3.82373 & 7.43136 & 0. & -0.0068586 & 0. & 3.54526 & 0. & 3.3935 & 1.7193 \\
176.488 & 32.4122 & 0.0776154 & -62.9606 & 0.177886 & -12.346 & 26.5689 & 7.04326 & 0. & -0.0170918 & 0. & -5.2778 & 0. & 1.56207 & 1.18349 \\
-21.5227 & -39.0633 & -0.494407 & 121.758 & -0.355958 & 0.576694 & -5.67678 & -17.58 & 0. & -0.00940279 & 0. & -3.59933 & 0. & -4.92509 & -4.18857 \\
22.4313 & 97.9958 & 1.19706 & -159.224 & 0.120322 & 24.8872 & 3.17625 & 23.7835 & 0. & -0.11519 & 0. & 5.8433 & 0. & 1.73174 & 2.14857 \\
\end{pmatrix}
\tag{27}
$$

変数 t，ev の積演算を表す変換行列も同様に計算することができる。変換行列 $m_t$ の固有ベクトル$v_\xi$を求め，関係式

$$m_t \cdot v_\xi = \xi_t \cdot v_\xi \tag{28}$$

$$m_{ev} \cdot v_\xi = \xi_{ev} \cdot v_\xi \tag{29}$$

から t, ev の値を固有値 $\xi_t$, $\xi_{ev}$として求めることができる。このとき，実数解は，表 2 に

示した多項式の三角化解法によって得られたものと同等である。

## 4. まとめ

以下，本手法の効果を順に挙げる。

(利点 1) 本手法では，基礎方程式系であるハートリー・フォック・ロータン方程式及び拘束条件・最適化条件を多項式系で表すため，電子状態計算と付加条件とがシームレスに結合される。従来法では，固有値計算のループ，自己無撞着計算のループ，最適化計算のループが入れ子になっており複雑である。一方，本手法においては，これらの入れ子になったループが，連立多変数多項式系の根の探索という単一のフローに統合されており，数値計算時に見通しのよいものであり，手短に必要な情報に到達することが可能である。従来法で行われたような，原子核・波動関数の自由度に関する独立の緩和計算を交互に反復する必要がない。本法では原子核と波動関数のダイナミクスが強く結合するような物理現象も、断熱近似の域を超えて、取り扱うことができるであろう。

(利点 2)
本手法においては基礎方程式系である HFR 方程式及び拘束条件・最適化条件を多項式系で表すもう一つ利点は次の通りである。多項式系は，未知変数間の関係式を与えるものである。したがって，多項式系によって記述される未知変数を求めたい場合，計算の入力・出力に，それぞれ，適切な変数を振り分けることができる。すなわち，入力値は構造であり，出力は電子状態である，という従来法の逆問題の枠組みに限定されない。したがって，従来法でいう順問題・最適化問題・逆問題の区別が解消され，すべて順問題として統一的に解くことができる。逆問題を扱う場合には，問題の適切性を議論しなくてはならない。本方法では，基礎方程式系をグレブナー基底に変換した段階で逆問題の適切性，すなわち解の存在の検証が可能である。代数学のイデアル理論に基づき，グレブナー基底の零値を与える解集合が存在するか否かが判る。解集合が存在する場合，それは孤立点か，あるいは，一次元以上の次元を持つ集合であるかがわかる。解集合が孤立点の場合，複素数あるいは実数の範囲で，幾つ解があるかも判定できる。

(利点 3) 本手法を自己無撞着場計算に適用する利点は次のようなものである。三角化された多項式系を一変数ずつ数値的に求めることによって計算できる。収束性に問題の発生することの多い，多次元行列に対する固有値計算も，平均場の逐次近似計算も不要である。

(利点 4) 本手法によって，最適化計算する場合の利点として，次の点も挙げることができる。例として，原子構造の最適化を考える。変数消去の順番を工夫し，原子座標のみの多項式を得ることができる。この多項式の根が安定構造である。すなわち，SCF 計算を行わずに，最適な原子配置を求めることができるのである。他のパラメータを最適化する際も同様の利点がある。

本研究は「分子積分の多項式近似」「分子軌道代数方程式」という概念が現実の系の電子状態計算に適用可能であるのみならず、既存手法では不可能あるいは困難が大きい問題も視野に収めうるものを示すものである。ただし、現状では、必ずしも精度の良い計算ではない。この理由の一つは、計算機性能の制約である。記号処理の計算負荷が大きく、このために、近似多項式の次数を小さくし、また、数値係数を低精度で有理数近似を行わざるを得ない。さらに、記号処理上の原理的な問題もある。式(14)に見るように、出発点とした多項式連立方程式系の数値係数の大きさはほぼ揃っている。しかし、これを数式処理し、グレブナー基底を生成すると、数値係数の大きさは極めて不揃いなものになる。この種の、

係数が極端に膨れ上がる事態は、計算コストを増大させるばかりでなく数値解上の精度悪化の原因となるが、現時点では有効な対策は少ない。しかしながら、今後の計算機性能の向上と数式処理技術の改良によって、本手法における十分な精度の達成と大規模計算への適用が可能になるであろう。

A proposal to first principles electronic structure calculation:
Symbolic-Numeric method

Akihito Hisai Kikuchi
CANON INC.
30-2, Shimomaruko 3-chome, Ohta-Ku, Tokyo 146-8501-Japan
E-mail:kikuchi.akihito@canon.co.jp

**Abstract**

This study proposes an approach toward the first principles electronic structure calculation with the aid of symbolic-numeric solving. The symbolic computation enables us to express the Hartree-Fock-Roothaan equation in an analytic form and approximate it as a set of polynomial equations. By use of the Gröbner bases technique, the polynomial equations are transformed into other ones which have identical roots. The converted equations take more convenient forms which will simplify numerical procedures, from which we can derive necessary physical properties in order, in an "a la carte" way. This method enables us to solve the electronic structure calculation, the optimization of any kind, or the inverse problem as a forward problem in a unified way, in which there is no need for iterative self-consistent procedures with trials and errors.

1. **Introduction**

The basic equation of the first principle electronic structure calculations is the Hartree-Fock or the Kohn-Sham equation, derived from the minimum condition of the energy functional in the electron-nuclei system. They are expressed as Schrodinger-type equations as follows [1-3].

$$\left( -\frac{1}{2}\Delta + \sum_a \frac{Z_a}{|r - R_a|} + \int dr' \frac{\rho(r')}{|r - r'|} + V^{exc} \right) \phi_i(r) = E_i \phi_i(r) \tag{1}$$

The second term in the parse of the left side is the potential from nuclei with charge Za. The third term is the Coulomb potential generated by the charge distribution ρ. The forth term $V^{exc}$ is the quantum dynamical interaction operating in the many-electron system, called the "exchange and correlation". To obtain the numerical solution, we expand the wavefunction by a certain bases set. The numerical procedure is converted into the following eigenvalue problem of matrices, if the wavefunctions are expressed as linear combinations of localized base (Hartree-Fock-Roothaan equation, hereafter abbreviated as HFR equation).

$$\hat{H}(\{R\},\{\alpha\},\{Q\},\{C\})C(\{R\},\{\alpha\},\{Q\}) = \quad (2)$$
$$\hat{S}(\{R\},\{\alpha\},\{Q\})C(\{R\},\{\alpha\},\{Q\})e(\{R\},\{\alpha\},\{Q\})$$

In the above, $\hat{H}$ is the Hamiltonian matrix, $\hat{S}$ is the overlap one, $C$ is the wavefunction (the coefficients of the linear combination), and $e$ is the eigenvalue. The variables {R} are the positions of the nuclei, {α} are the orbital exponents which describe the special expansion of the localized bases function, and {Q} are the quantum numbers. The localized atomic bases are parameterized by them. The HFR equation can be expressed by multi-valuable analytic functions, whose variables are {R}, {α} and {Q}. It contains transcendental functions of several kinds. This is because we conventionally adopt analytical base, such as STO (Slater type orbital) or GTO (Gaussian type orbital), to construct one-electron and two-electron integrals. The use of analytic base is effective in numerical treatment, but arises some difficulty in the mathematical operations to the complicated analytic expression. We, however, can approximate the equation by polynomials with respect to these variables and obtain much simpler expressions (Weierstrass approximation theorem). For this purpose, it is sufficient for us to perform the Taylor expansion. The equation becomes the set of algebraic polynomial equations expressed by atomic coordinates, orbital exponents, and quantum numbers. The concrete expression can be derived through symbolic manipulation [4-9] by means of computer algebra systems [10, 11]. The concept of the polynomial approximation to the HFR equation in the molecular orbital method is proposed by Yasui [6-9], in which this approximated equation is designated as the "Algebraic molecular orbital equation". Based on this, we will able to unravel the relationship among parameters and clarify their dependence.(See the appendix.)

The conventional method of the electronic structure calculations is a "forward problem". We suppose the structural data of the material, execute the electronic structure calculations, optimize the structure so that the energy functional will be minimized, and evaluate electronic properties in the stable structure. The conventional method has some inefficiency. For example, in the first principles molecular dynamics, the nuclei are determined by the classical Newtonian equation, while the wavefunctions are ruled by quantum dynamics. The foundation for this treatment is so-called "the adiabatic approximation", which enables us to separate the dynamics of the nuclei and wavefunctions into two independent models. This conventional method iterates two alternative computational phases, one of which are the optimization for the wavefunctions and the other for the positions of the nuclei. It is believed that this way is numerically stable. But, in view of effectiveness of the optimization, this may be a lengthy and roundabout one. Owing to the separation of the degrees of freedom of wavefunctions and nuclei, it is difficult for the conventional method to coop with cases where the dynamics of nuclei and wavefunctions are strongly coupled with each other. Meanwhile, the "inverse problem" will be to search the material structure which shows the desirable electronic properties. In the treatment of the inverse problems by means of the conventional method, we must go with trial and error. At first we suppose the material structure to evaluate the electronic properties, and, by adjusting the structure, we search the direction in which the desired properties will be obtained. We have to solve forward problems repeatedly to obtain the solution of the inverse problems. The reason to this is as follows. In the conventional methods, the computation has the fixed order of numerical procedures, consisted from the eigenvalue solution, the self-consistent-field calculation and the relaxation of atomic structure, which is implemented as nested loops of independent phases of the optimizations. The unknown parameters are computed from inner loops to outer ones in order. The conventional method is obliged to determine unknown variables in a fixed order in any cases.

## 2. Symbolic-numeric ab-initio molecular dynamics and molecular orbital method

With the view of these circumstances, we propose the following method.

It is summarized as follows. "At first, HFR equation is approximated as a set of multi-variable polynomial equations, and through the symbolic computation, it is transformed into a certain form more convenient for the numerical treatment. The eigenstates are evaluated by the root finding by means of symbolic-numeric procedure. "

The question in the present work is how we can obtain the numerical solutions of the equations after the polynomial approximation and derive the significant information. In general, when we try to solve a set of polynomial equations with finite numbers of solutions, we numerically compute them using the floating-point approximation. However, under the influence of numerical errors, the genuine numerical methods often become unstable and it is difficult to predict the tendency of the drifting. Thus, to avoid numerical instabilities, the several types of techniques, so-called "symbolic-numeric solving", are proposed. In them, the symbolic manipulation is applied as a preconditioning to the set of equations to be solved. The equations are transformed into the forms which have the same roots, to which the numerical computation will be easy and stable. From the form of the transformed equation, the character of the solution, such as, the existence and the geometrical structure, can be determined. Then the solving process is passed over to the numerical one. For the mathematical background, see ref. [12-15]. A review of application of symbolic computations in the field of the computational chemistry is given in ref .16.

In order to solve the HFR equation, the present method makes use of the several ways of the symbolic-numeric solving, by which, the HFR equation, approximated as a set of polynomial equations, is transformed into one which has the same roots. As a strategy, the algorithm of the "decomposition of polynomial equations into triangular sets" is applied [12, 13].   In this algorithm, the following transformations are applied.

The starting equations $f_1, \dots, f_n$

$$f_1(x_1, x_2, \ldots, x_n) = 0 \qquad (3)$$
$$f_2(x_1, x_2, \ldots, x_n) = 0$$
$$\vdots$$
$$f_n(x_1, x_2, \ldots, x_n) = 0$$

→ Gröbner bases with lexicographic order of $f_1, \ldots, f_n$  $\{g_i\}$

$$g_1(x_1) = 0 \qquad (4)$$
$$\vdots$$
$$g_{2\_1}(x_1, x_2) = 0$$
$$\vdots$$
$$g_{2\_m(2)}(x_1, x_2) = 0$$
$$g_{3\_1}(x_1, x_2, x_3) = 0$$
$$\vdots$$
$$g_{n\_1}(x_1, \ldots, x_n) = 0$$
$$\vdots$$
$$g_{n\_m(n)}(x_1, \ldots, x_n) = 0$$

→ Triangular sets of polynomials  $\{t_i\}$

$$t_1(x_1) = 0 \qquad (5)$$
$$t_2(x_1, x_2) = 0$$
$$\vdots$$
$$t_n(x_1, x_2, \ldots, x_n) = 0$$

The algorithm in ref .12 and 13, at first, generates the Gröbner bases $\{g_i\}$ with the lexicographic monomial ordering from the starting set of equations $f_1, \ldots, f_n$. The generated Gröbner bases have roots identical to those of $f_1, \ldots, f_n$, and take forms which guarantee an easier numerical solving. The Gröbner bases are a set of polynomials in which the number of unknowns of each entry increases in order, from polynomials with fewer valuables to ones with more. However, the total number of polynomials in the Gröbner bases may grow more than that of the starting polynomials. Though we can search the root at this stage, we furthermore apply the decomposition to the Gröbner bases. By means of it, we obtain several "triangular" systems of equations $\{t_1, \ldots, t_n\}$, the first entry of which has one unknown $x_1$, the second has two unknowns $x_1, x_2$, the third three unknowns, etc, until, the last n-th has n-unknowns $x_1, x_2, \ldots, x_n$ by turns.  In order to obtain all roots of the starting equations $f_1, \ldots, f_n$, we may need to construct several sets of triangular set of equations, all of which can be generated by the algorithm in ref .12 or 13. Single triangular system has fewer numbers of entries than that in the Gröbner bases before the transformation. This makes the numerical procedure easier. Once we can obtain the triangular systems of the equations, we can evaluate each unknown one by one. In the numerical solution, only a Quasi-Newton-like method, or its kindred for one variable, is necessary.

We can execute another type of the symbolic numerical solving. The foundation to this is the theorem of Stickelberger. As above, we regard the HFR equations as the set of polynomial equations expressed by unknowns $X_1, X_m$. The set of polynomial equations constructs a zero-dimensional ideal I in the polynomial ring $R=k[X_1,…,X_m]$, whose zeros corresponds to a residue ring $A=R/I$. Here k is the coefficient field, which, in our cases, is the rational number field or the real number field. The ring A is a finite dimensional vector space over k, whose bases are expressed by monomials of $X_1, X_2, …, X_m$. Thus, the multiplication by $X_1, X_2, …, X_m$ on each base results in the linear combination of the bases in A. The product operations by $X_1, X_2, …, X_m$ are expressed as a linear transformation matrix $m_h$ ( $h = X_1, X_2, …, X_m$). The bases in the residue ring $A=R/I$ and the transformation matrices are obtained by means of Gröbner bases technique. The theorem of Stickelberger asserts that there is a one-to-one correspondence between an eigenvector $v_\xi$ of the matrix $m_h$ and the zero point $\xi = (\xi_1, …, \xi_m)$ of the ideal I. The correspondence is given by (6).

$$m_{X_i} \cdot v_\xi = \xi_i \cdot v_\xi \qquad (6)$$

The eigenvector $v_\xi$ is common with all of $m_{X_i}$. (For details, see ref.14, p101-130, "From Enumerative Geometry to Solving Systems of Polynomial Equations", by Frank Sottile.) The numerical calculations for the zeros $\xi = (\xi_1, …, \xi_m)$ are executed as follows. We choose one $X_i$ and prepare $m_{X_i}$. The above secular equation gives us the eigenvector $v_\xi$. If we multiply $m_{X_j}$ ($j \neq i$) with $v_\xi$, we can evaluate $\xi_j$ ($j \neq i$). Thus all values of $\xi = (\xi_1, …, \xi_m)$ can be obtained.

The energy functional, the normalization conditions for wavefunctions, and the HFR equation are polynomials with respect to LCAO(Linear Combination of Atomic Orbitals) coefficients and eigenvalues. Those equations are constructed from molecular integrals, which are in general expressed as analytic functions of the included parameters, accordingly not being polynomials. Thus, the molecular integrals are replaced by approximations of polynomials of the included parameters. By means of this, we can construct a set of polynomial equations including not only wavefunctions and eigenvalues but also the parameters for molecular integrals, i.e. the molecular orbital algebraic equation. If extra constraint conditions should be cast upon the HFR equation, we can prepare the polynomial equations for the constraints and add them into the set of polynomial equations. If the numerical coefficients are rationalized, we can avoid the lowering of the precision through the symbolic manipulation by means of arbitrary precision calculations of rational numbers.

The one of the merits in this treatment is as follows. In the conventional method, the input data is the atomic structure and the output if the electronic structure. By contrast, in the present method, the input data are not confined to the atomic structures. We can select arbitrary parameters in the HFR equation and its constraints and set them as the input. If the problem to be solved is properly established, we can compute other unknown variables properly. As to the properness of the problem, i.e., the existence of the roots of the set of polynomial equations, it can be judged from the ideal theory in mathematics on the condition whether its Gröbner bases have zero points set or not.

The task flow is listed as follows.

1. Compute the analytic formula of the energy functional, whose variables are eigenvalues, LCAO coefficients in wavefunctions, atomic coordinates, and orbital exponents in molecular integrals and so on.
2. Express the constraint conditions as analytic formulas. By rationalizing the numerical coefficients, we can prevent the lowering of the precision through the afterward symbolic manipulation.
3. Those analytic expressions are polynomials with respect to LCAO coefficients, while the expressions are not polynomials with respect to other parameters. Molecular integrals are replaced by approximating polynomials with respect to the included parameters. For example, by choosing a certain point of a parameter and applying the Taylor expansion around it, we can obtain the polynomial approximation.
4. Prepare the set of the equations to be solved, by way of the minimization of the energy functional approximated by the polynomials.
5. By means of the symbolic manipulation, the above equations are transformed into other ones having the same roots. The initial equations are, at first, transformed into the Gröbner bases, by which we can check the existence of the solution. If the solutions are the set of isolated points, we can decompose the Gröbner bases into the triangular expression. There is an alternative way: by means of Stickelberger's theorem, the search for the solution is replaced by an eigenvalue problem.
6. By numerically solving the equations, the roots are computed and afford us the electronic structure and the useful information.

### 3. The demonstrations of the calculations

Several examples are demonstrated in this section. The units are given in atomic units.

At first, we will show the possibility of the SCF (Self-Consistent Field) calculations. As an object, we choose the hydrogen molecule. Though we only show the examples of $H_2$ here, which is the simplest molecule, the applications of the present method are not limited to two-electron or two-atomic systems. The reason why we choose $H_2$ is as follows. Though this system is simple, it includes all kinds of quantum interactions operating in realistic materials and can be assumed as a miniature of general many-electron and polyatomic systems. The molecular integrals needed here are generated by STO (Slater Type Orbital) base. The energy functional is the analytic equation of the one centered molecular integrals at hydrogens A and B, the two-centered molecular integrals, and the coefficients of the wavefunctions (a, b, c, d). As the expression of the interatomic distance R, the molecular integrals contain transcendental functions. One of the two-centered molecular integrals is shown in (7). This is the two electron repulsion between 1s orbitals and classified as "Coulombic type".

This is denoted as $[1s(A)1s(A)||1s(B)1s(B)] = \iint drdr' \frac{\phi^{1s}(r-R_A)\phi^{1s}(r-R_A)\phi^{1s}(r'-R_B)\phi^{1s}(r'-R_B)}{|r-r'|}$,

where 1s(A) and 1s(B) mean the atomic orbitals centered on atoms A and B. In addition to this, there are other repulsion integrals, denoted as the "exchange type" $[1s(A)1s(B)||1s(A)1s(B)]$ and the "hybrid type" $[1s(A)1s(A)||1s(A)1s(B)]$. In general, the molecular integrals have more complicated expressions than (7). If STO is used, the integrals include exponentials and the exponential integration, and, as for the exchange type, it is expressed by infinite series. Though these are complicated analytic formulas, we can treat them easily in the symbolic processing, by approximating them as finite degree polynomials by means of the Taylor expansion. It is noted here that the STO base can describe the physical property of the localized atomic wavefunction more precisely than by GTO (Gaussian Type Orbital) base, both in the neighborhood of the nucleus and in the remote region from it. Thus the STO base becomes more advantageous for the purpose of expressing the molecular equations as the polynomials of the atomic coordinates. This is the reason why STO is adopted here. However, the following recipes are also applicable to the GTO calculations, and possibly, to semi-empirical calculations, such as AM1 (Austine Model 1) [17] and PM3 (Parameterized Model number 3) [18], or tight-binding model, where the matrix elements are given as analytic formulas.

$$[1s(A)1s(A)|1s(B)1s(B)] = \qquad (7)$$

$$\frac{64\, za^{3/2}\, zb^{3/2}\, zc^{3/2}\, zd^{3/2}}{R\,(za+zb)^3\,(zc+zd)^3\, E^R} - \frac{32\, za^{3/2}\, zb^{3/2}\, (za+zb)\, zc^{3/2}\, zd^{3/2}}{E^{R(zc+zd)}\,(za+zb-zc-zd)^2\,(zc+zd)^2\,(za+zb+zc+zd)^2}$$

$$- \frac{32\, za^{3/2}\, zb^{3/2}\, zc^{3/2}\, zd^{3/2}\,(zc+zd)}{E^{R(za+zb)}\,(za+zb)^2\,(za+zb-zc-zd)^2\,(za+zb+zc+zd)^2}$$

$$+ \frac{64\, za^{3/2}\, zb^{3/2}\, zc^{3/2}\, zd^{3/2}\,(zc+zd)\,(3\, za^2 + 6\, za\, zb + 3\, zb^2 - zc^2 - 2\, zc\, zd - zd^2)}{E^{R(za+zb)}\, R\,(za+zb)^3\,(za+zb-zc-zd)^3\,(za+zb+zc+zd)^3}$$

$$+ \frac{64\, za^{3/2}\, zb^{3/2}\,(za+zb)\, zc^{3/2}\, zd^{3/2}\,(-za^2 - 2\, za\, zb - zb^2 + 3\, zc^2 + 6\, zc\, zd + 3\, zd^2)}{E^{R(zc+zd)}\, R\,(za+zb-zc-zd)^3\,(zc+zd)^3\,(za+zb+zc+zd)^3}$$

As an example of a forward problem in the first principles molecular dynamics, the optimization of the structure (the distance between two hydrogens) and the UHF electronic structure calculation are simultaneously executed.

Here, the position of hydrogens A and B is denoted as $R_A$ and $R_B$. The interatomic distance is $r = R_A - R_B$. We use the notation, such as $x_A = |x - R_A|$ and $x_B = |x - R_B|$.

We execute the UHF (Unrestricted Hartree-Fock) calculations in which the trial wavefunctions for up- and down-spins are defined as in (8) and (9). The corresponding eigenvalues are denoted as ev and ew. It is noted here: these trial functions with the bases of the orbital exponent 1 are too primitive to assure good agreements with experiments: for accuracy, we must optimize the orbital exponent to a suitable value. It is only to reduce the computational cost in the symbolic computation why we adopt such primitive trial functions.

$$\phi_{up}(x) = (a\exp(-x_A) + b\exp(-x_B))/\sqrt{\pi} \tag{8}$$

$$\phi_{down}(x) = (c\exp(-x_A) + d\exp(-x_B))/\sqrt{\pi} \tag{9}$$

The energy functional is transformed into the polynomial form by way of the fourth order Taylor expansion of the interatomic distance r, centered at the position of $R_0=7/5$ atomic unit. The functional is generated in a standard way of molecular orbital theory, which is the total energy of the electron-nuclei system (given in the atomic units) with the constraint condition of the ortho-normality of the wavefunctions. The Lagrange multipliers are eigenvalues. The coefficients of real numbers are truncated to third decimal places and approximated as rational numbers, as is shown in (10). (a,b) are the LCAO coefficients for the up-spin electron,(c,d) are those for the down-spin electron, ev is the eigenvalues of the up-spin electron, ew is that of the down-spin electron, r is the interatomic distance.

$$\Omega[\{\phi_i(\xi); \text{ the occupied orbital } i, \xi \equiv (r,\sigma)\}] \tag{10}$$

$$= \sum_i \int d\xi\, \phi_i(\xi)\left(-\frac{1}{2}\nabla^2 + \sum_a \frac{Z_a}{|r-R_a|}\right)\phi_i(\xi)$$

$$+ \frac{1}{2}\sum_{i,j} \iint d\xi d\xi'\, \frac{\phi_i(\xi)\phi_i(\xi)\phi_j(\xi')\phi_j(\xi')}{|r-r'|}$$

$$+ \frac{1}{2}\sum_{i,j} \iint d\xi d\xi'\, \frac{\phi_i(\xi)\phi_j(\xi)\phi_j(\xi')\phi_i(\xi')}{|r-r'|}$$

$$+ \frac{1}{2}\sum_{a,b(a\neq b)} \frac{Z_a Z_b}{|R_a-R_b|}$$

$$+ \sum_{i,j} \lambda_{ij}\left(\int d\xi \phi_i(\xi)\phi_j(\xi) - \delta_{ij}\right)$$

⇓

Ω＝(3571 - 1580*a^2 - 3075*a*b - 1580*b^2 - 1580*c^2 + 625*a^2*c^2 + 1243*a*b*c^2 + 620*b^2*c^2 - 3075*c*d + 1243*a^2*c*d + 2506*a*b*c*d + 1243*b^2*c*d - 1580*d^2 + 620*a^2*d^2 + 1243*a*b*d^2 + 625*b^2*d^2 + 1000*ev - 1000*a^2*ev - 1986*a*b*ev - 1000*b^2*ev + 1000*ew - 1000*c^2*ew - 1986*c*d*ew - 1000*d^2*ew - 5102*r + 332*a^2*r + 284*a*b*r + 332*b^2*r + 332*c^2*r + 43*a*b*c^2*r + 20*b^2*c^2*r + 284*c*d*r + 43*a^2*c*d*r + 80*a*b*c*d*r + 43*b^2*c*d*r + 332*d^2*r + 20*a^2*d^2*r + 43*a*b*d^2*r - 63*a*b*ev*r - 63*c*d*ew*r + 3644*r^2 + 75*a^2*r^2 + 724*a*b*r^2 + 75*b^2*r^2 + 75*c^2*r^2 - 401*a*b*c^2*r^2 - 124*b^2*c^2*r^2 +

724*c*d*r^2 - 401*a^2*c*d*r^2 - 1372*a*b*c*d*r^2 - 401*b^2*c*d*r^2 + 75*d^2*r^2 - 124*a^2*d^2*r^2 - 401*a*b*d^2*r^2 + 458*a*b*ev*r^2 + 458*c*d*ew*r^2 - 1301*r^3 - 69*a^2*r^3 - 303*a*b*r^3 - 69*b^2*r^3 - 69*c^2*r^3 + 146*a*b*c^2*r^3 + 42*b^2*c^2*r^3 - 303*c*d*r^3 + 146*a^2*c*d*r^3 + 618*a*b*c*d*r^3 + 146*b^2*c*d*r^3 - 69*d^2*r^3 + 42*a^2*d^2*r^3 + 146*a*b*d^2*r^3 - 139*a*b*ev*r^3 - 139*c*d*ew*r^3 + 185*r^4 + 12*a^2*r^4 + 39*a*b*r^4 + 12*b^2*r^4 + 12*c^2*r^4 - 17*a*b*c^2*r^4 - 4*b^2*c^2*r^4 + 39*c*d*r^4 - 17*a^2*c*d*r^4 - 86*a*b*c*d*r^4 - 17*b^2*c*d*r^4 + 12*d^2*r^4 - 4*a^2*d^2*r^4 - 17*a*b*d^2*r^4 + 13*a*b*ev*r^4 + 13*c*d*ew*r^4)/1000

The precision of this approximation should be checked at first. The values of the original energy functional and the polynomial approximation at $a = b = c = d = 1$ and $ev = ew = 0$ are plotted in Fig.2 as the function of R. It shows sufficient agreement in the range of r=1～2 atomic unit. However, if the r goes out of this region, the polynomial approximation is not appropriate and we must try another center of the expansion $R_0$. By increasing the maximum degree of the Taylor expansion, we can enlarge the range of r where the polynomial approximation is valid.

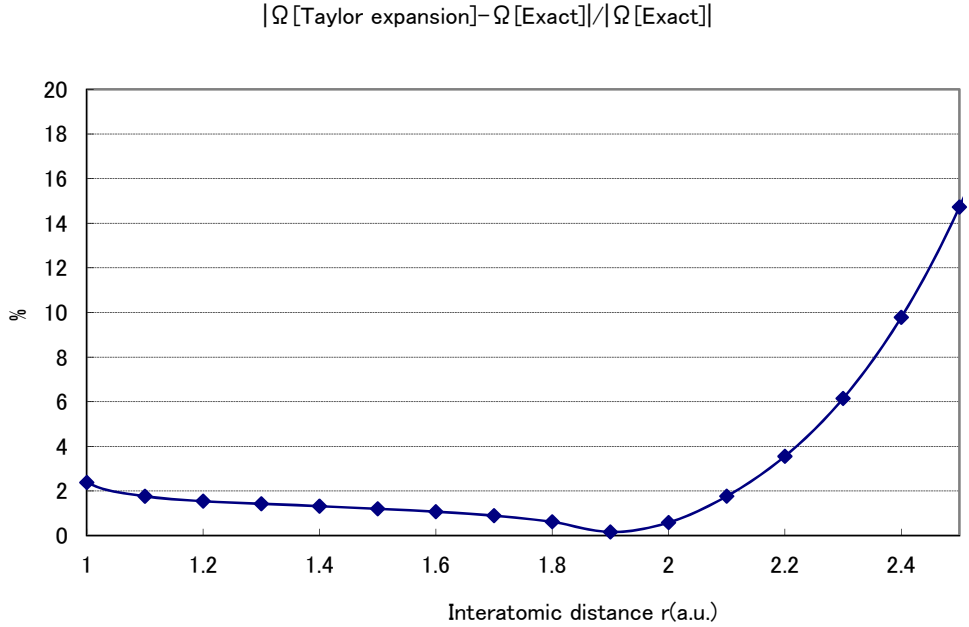

Figure 2. This figure shows the deviation between the exact energy functional $\Omega$ and its polynomial approximation, $|\Omega(\text{Taylor expansion}) - \Omega(\text{Exact})|/|\Omega(\text{Exact})|$ v.s. the interatomic distance r, given in percentage.

Here, we make use of the symmetry in $H_2$, and we express the wavefunctions as the linear combination of the symmetric and asymmetric ones, as in (11) and (12).

$$\phi_{up}(x) = t(\exp(-x_A) + \exp(-x_B))/\sqrt{\pi} + s(\exp(-x_A) - \exp(-x_B))/\sqrt{\pi} \quad (11)$$

$$\phi_{down}(x) = u(\exp(-x_A) + \exp(-x_B))/\sqrt{\pi} + v(\exp(-x_A) - \exp(-x_B))/\sqrt{\pi} \quad (12)$$

This is the transformation by (13).

$$a = t+s, b = t-s, c = u+v, d = u-v \quad (13)$$

Then the HFR equation is given by the set of equations in (14), where (t,s) is the LCAO coefficient for up-spin, (u,v) is that for down spin, ev is the eigenvalue for up spin, ew is that for down spin, and r is the interatomic distance.

① $\dfrac{\partial \Omega}{\partial a} + \dfrac{\partial \Omega}{\partial b} = 0 \rightarrow$ (14)

32*s*u*v*r^4-336*s*u*v*r^3+992*s*u*v*r^2-160*s*u*v*r+40*s*u*v-324*t*u^2*r^4+2572*t*u^2*r^3-6448*t*u^2*r^2+584*t*u^2*r+19936*t*u^2+156*t*v^2*r^4-1068*t*v^2*r^3+2248*t*v^2*r^2-80*t*v^2*r-32*t*v^2+26*t*ev*r^4-278*t*ev*r^3+916*t*ev*r^2-126*t*ev*r-7972*t*ev+126*t*r^4-882*t*r^3+1748*t*r^2+1896*t*r-12470*t=0

② $\dfrac{\partial \Omega}{\partial a} - \dfrac{\partial \Omega}{\partial b} = 0$

→156*s*u^2*r^4-1068*s*u^2*r^3+2248*s*u^2*r^2-80*s*u^2*r-32*s*u^2-52*s*v^2*r^4+236*s*v^2*r^3-32*s*v^2*r^2-104*s*v^2*r+48*s*v^2-26*s*ev*r^4+278*s*ev*r^3-916*s*ev*r^2+126*s*ev*r-28*s*ev-30*s*r^4+330*s*r^3-1148*s*r^2+760*s*r-170*s+32*t*u*v*r^4-336*t*u*v*r^3+992*t*u*v*r^2-160*t*u*v*r+40*t*u*v=0

③ $\dfrac{\partial \Omega}{\partial c} + \dfrac{\partial \Omega}{\partial d} = 0$

→156*s^2*u*r^4-1068*s^2*u*r^3+2248*s^2*u*r^2-80*s^2*u*r-32*s^2*u+32**t*v*r^4-336*s*t*v*r^3+992*s*t*v*r^2-160*s*t*v*r+40*s*t*v-324*t^2*u*r^4+2572*t^2*u*r^3-6448*t^2*u*r^2+584*t^2*u*r+19936*t^2*u+26*u*ew*r^4-278*u*ew*r^3+916*u*ew*r^2-126*u*ew*r-7972*u*ew+126*u*r^4-882*u*r^3+1748*u*r^2+1896*u*r-12470*u=0

④ $\dfrac{\partial \Omega}{\partial c} - \dfrac{\partial \Omega}{\partial d} = 0$

→-52*s^2*v*r^4+236*s^2*v*r^3-32*s^2*v*r^2-104*s^2*v*r+48*s^2*v+32*s*t*u*r^4-336*s*t*u*r^3+992*s*t*u*r^2-160*s*t*u*r+40*s*t*u+156*t^2*v*r^4-1068*t^2*v*r^3+2248*t^2*v*r^2-80*t^2*v*r-32*t^2*v-26*v*ew*r^

4+278*v*ew*r^3-916*v*ew*r^2+126*v*ew*r-28*v*ew-30*v*r^4+330*v*r^3-1148*v*r^2+760*v*r-170*v=0

⑤ $\dfrac{\partial \Omega}{\partial (ev)} = \langle \phi_{up} | \phi_{up} \rangle - 1 = 0$

→-13*s^2*r^4+139*s^2*r^3-458*s^2*r^2+63*s^2*r-14*s^2+13*t^2*r^4-139*t^2*r^3+458*t^2*r^2-63*t^2*r-3986*t^2+1000=0

⑥ $\dfrac{\partial \Omega}{\partial (ew)} = \langle \phi_{down} | \phi_{down} \rangle - 1 = 0$

→13*u^2*r^4-139*u^2*r^3+458*u^2*r^2-63*u^2*r-3986*u^2-13*v^2*r^4+139*v^2*r^3-458*v^2*r^2+63*v^2*r-14*v^2+1000=0

⑦ $\dfrac{\partial \Omega}{\partial r} = 0$

→312*s^2*u^2*r^3-1602*s^2*u^2*r^2+2248*s^2*u^2*r-40*s^2*u^2-104*s^2*v^2*r^3+354*s^2*v^2*r^2-32*s^2*v^2*r-52*s^2*v^2-52*s^2*ev*r^3+417*s^2*ev*r^2-916*s^2*ev*r+63*s^2*ev-60*s^2*r^3+495*s^2*r^2-1148*s^2*r+380*s^2+128*s*t*u*v*r^3-1008*s*t*u*v*r^2+1984*s*t*u*v*r-160*s*t*u*v-648*t^2*u^2*r^3+3858*t^2*u^2*r^2-6448*t^2*u^2*r+292*t^2*u^2+312*t^2*v^2*r^3-1602*t^2*v^2*r^2+2248*t^2*v^2*r-40*t^2*v^2+52*t^2*ev*r^3-417*t^2*ev*r^2+916*t^2*ev*r-63*t^2*ev+252*t^2*r^3-1323*t^2*r^2+1748*t^2*r+948*t^2+52*u^2*ew*r^3-417*u^2*ew*r^2+916*u^2*ew*r-63*u^2*ew+252*u^2*r^3-1323*u^2*r^2+1748*u^2*r+948*u^2-52*v^2*ew*r^3+417*v^2*ew*r^2-916*v^2*ew*r+63*v^2*ew-60*v^2*r^3+495*v^2*r^2-1148*v^2*r+380*v^2+740*r^3-3903*r^2+7288*r-5102=0

At first, we assume r=7/5 to see the possibility of actual first principles calculations. The entry as is shown by (15) in (14)

$$\dfrac{\partial \Omega}{\partial r} = 0 \qquad (15)$$

is replaced by (16).

$$5r - 7 = 0 \qquad (16)$$

The lexicographic order Gröbner bases are shown in (17) with the monomial ordering of s<t<u<v<ev<ew<r. (Hereafter this monomial ordering is kept in the following computations.) In the entries of the Gröbner bases, those variables show themselves in the reverse order of r, ew, ev, v, u, t, s, by turns. The relationships among those variables, which are ambiguous in the expression of HFR equation, may be extracted there.

J[1]=r-1.4 (17)

J[2]=0.0000000000000004492367995028017983415375254583712005086 9800531719*ew^6+0.00000000000000083784847684948179918230616638 79813241022968255965*ew^5+0.0000000000000005522431406589942780 2789765894852689587104815167352*ew^4+0.00000000000000014691161 15773407851793157859874999641802949934 6849*ew^3+0.0000000000 000011979351185242681597299140874114661346717404726834*ew^2-0.00 000000000000000839869963562374452466573406585698958493547765 49449*ew-0.00000000000000000003723040905315595026398352218369 1557632339096670694

J[3]=0.000000001396287653570571574662207138195309771989211807 0071*ev+0.0000003217365693623504793206423800429601988107513122 5526*ew^5+0.0000003329165247773600520718203430909808162869659 4819331*ew^4+0.00000006589110323648302126095620051055593446942 5614980251*ew^3-0.000000014502194796324730495895492585449540487 87083075 49*ew^2-0.000000025896639470303145276823808149265037 63029302261 6944*ew+0.0000000005081975132258812443523091396 9686213591494118136

J[4]=0.00353374212492811163514143308777528431255435619356 16*v*ew^4+0.002180171063553251204119252395550142117611305286 8825*v*ew^3+0.00024683412076704120868 49546037854544501562730 166012*v*ew^2-0.00000145546516842582525115018507036947117736 79621730295*v*ew-0.0000007520336491246198410768915301515145 6273923670860722*v

J[5]=0.000000000000024691822639125973271970 7854366500885 8052 30838289*v^2-0.000000000047190217319792339322638580120 9311776724 87391351614*ew^5-0.000000000074233725861437012 5563724850891 3491 2758173742859754*ew^4-0.000000000037044 8756391838428180285450 40311021870857515 76865*ew^3-0.00000 0000055056937948011551503818 116670122630281056218 86299*ew^2 +0.0000000000000289247109748205722101525546739 21687498485750 622516*ew-0.000000000000048519950 0275950858789604761685291 06170995782184575

J[6]=0.000824445295339140048012174925203485761427508716 3634*u*ew^4+0.00154024712861099035299602345961334383656856 2434456*u*ew^3+0.00101861551014126740410975621397452429602 75920262626*u*ew^2+0.000273301471517934491088097554697358 4926704784843595*u*ew+0.0000231526498118394429461881625790 39171063101525686031*u

J[7]=0.15500652085373657278272155048840425269081666861422 *u*v*ew^2+0.09612448674592379845302508633008325276163013 6665135*u*v*ew+0.011178119900970398824611330746954877639528 183070683*u*v

J[8]=0.0000000000000246918226391259732719708 7854366500885 80523 0838289*u^2+0.00000000000066185419621956626251845869 357753738900

545315696192*ew^5+0.000000000010411459355962387028179380989969
086880336301691908*ew^4+0.000000000005195227570060962098498758
12168990888395938451519 3*ew^3+0.0000000000007218685315543857
62935403026499939648408016279623*ew^2-0.0000000000000040567605
78869777434741548572196422923655933 7475379*ew-0.0000000000000000
23368741490554450106147082788613043426219204168902

J[9]=0.06197091833163216567183870035153503888001513514567*t*ew^4
+0.07786967298515361066019891920879105862334446468288*t*ew^3+0.0
2952928754574380870962044724506192695086361951332 1*t*ew^2+0.0032
17837502146734951260237098620435482764135949341 1*t*ew+0.00004345
68530318230700128985173460828330264279573540 6*t

J[10]=0.05187413667297784999868866991024835274577850308566*t*v*ew
^3+0.03298141656889162536566431212120677368524682576014 2*t*v*ew^
2+0.0042447701517134049943862889216523608128280851170331*t*v*ew+
0.00005860046167413441588422093257956926678479916969229 5*t*v

J[11]=0.02179170335426535988026264693729198807682131368572 2*t*u*e
w^3+0.02704103778330774398428661678394321121208245598944 2*t*u*e
w^2+0.00996019886169329319975305970035709497670464004275 24*t*u*e
w+0.00097550648450924773054739454162691886820402290190827*t*u

J[12]=0.000000000000000246918226391259732719708785436650088580 52
30838289*t^2-0.0000000000019897101060519652107395125478208499820
923810270269*ew^5-0.0000000000017025570089756655535494287810528 3
795981196597736 02*ew^4-0.0000000000002348359220030419934874215 0
925412718812994059865964*ew^3+0.000000000001967372781280848501
57481822936643807630855116924 87*ew^2+0.00000000000001978607845 7
0238330525920261105821026866390535766*ew-0.0000000000000004421
9079290739404492269360637167466142801279295227

J[13]=0.05608483317512667868649481132510638542551836978360 4*s*ew
^2+0.034127439269225675895016122448171947840127214030889*s*ew-0.
00066282300441698232705078773127382486945402101120119*s-0.040893
88502024946597355319193847859030996007939963 8*t*u*v*ew-0.0290103
019717623826823207434166500720007104 82565746*t*u*v

J[14]=0.02179170335426535988026264693729198807682131368572 2*s*v*e
w-0.00041052862271488237533112188857801258113775891824163*s*v-t*u
*ew^2-0.915232433317809793085161470678692654025853166449 18*t*u*e
w-0.18279835739609284049315718860436264109118081404686*t*u

J[15]=0.08268962174313252600656152101084798951526898287462 1*s*u*e
w+0.05187413667297784999868866991024835274577850308566*s*u+t*v*e
w^2+0.45088158483424862414602508768019029048144708971225*t*v*ew
+0.00681768234810380611795226815290124500741917725349 97*t*v

J[16]=0.06197091833163216567183870035153503888001513514567*s*u*v
+7.54843402550665153673201812730961379367248454713989*t*ew^3+9.2

        9775966257943588582612339280270317033002044556 12*t*ew^2+2.933336
        9220816544967522410110109074685195419653 9922*t*ew+0.043698233453
        37914744233968162126461210128238625 2875*t

J[17]=0.15500652085373657278272155048840425269081666861422*s*t-u*v
        *ew-0.31006594502119821790385986369248662988955 02223605*u*v

J[18]=0.00000000000000246918226391259732719708 78543665008858052
        30838289*s^2+0.00000000014186637003179206604 9786475011937123318
        71881329982*ew^5+0.000000000121392348513937 0932693489804708248
        07184770749211 04*ew^4+0.0000000000016743805897305 1827655402125
        10939603762998421893102*ew^3-0.000000000014 02737183325013576178
        86142360602250481857722133 58*ew^2-0.00000000000001410747786486290
        63444287891909984518147600123 08984*ew-0.000000000000001865792945
        33155 971872568569090001959397434 54954377

The triangular decomposition to (17) is shown in (18), which involves five decomposed sets of equations. One decomposed set includes seven entries, into each of which, the seven variables is added one by one, with the order of r, ew, ev, v, u, t, s.

[1]:                                                                                                       (18)
  _[1]=r-1.4
  _[2]=0.08268962174313252600656152101084798951526 8982874621*ew
    +0.05187413667297784999868866991024835274577850308566
  _[3]=0.00000000139628765357057157466220713819530977198921 18070071*ev
    +0.00000000021872894505215127104699976329395744001395648217513
  _[4]=v
  _[5]=0.00000000000000246918226391259732719708785436650088 5805230838289*u^2
    -0.00000000000000070387276012025253831506975421079541060390364383305
  _[6]=t
  _[7]=0.00000000000000246918226391259732719708785436650 0885805230838289*s^2
      -0.000000000000005018614175944212997916038187913309155887 3472043949

[2]:
  _[1]=r-1.4
  _[2]=ew-0.01883875785389343111570679506560108729357957 3640499
  _[3]=ev-0.01883875785389343111570679506560108729357957 3640499
  _[4]=v^2-2.03250049593012100694952569568426974696181825868366
  _[5]=u
  _[6]=t
  _[7]=0.00000000000000246918226391259732719708785436650088580 5230838289*s^2
    -0.000000000000005018614175944212997916038187913309155887 3472043949

[3]:
  _[1]=r-1.4

_[2]=ew^2+0.62013189004239643580771972738497325977910044472l*ew
       +0.072113868754708844917382916961852739039135803532834
  _[3]=ev-ew
  _[4]=v^2-14.80995023428197133951905921507729315089177369253934*ew
       -5.777653339333545581487020716322560290468549512904ll
  _[5]=u^2+2.077131292262844160250221216181327899549542410218S4*ew
       +0.525266732317117545403256347138142609545167168240S5
  _[6]=t^2+2.077131292262844160250221216181327899549542410218S4*ew
       +0.525266732317117545403256347138142609545167168240S5
  _[7]=s+8.556707658075617499772076904610507582973719342317A6*t*u*v*ew
        +6.248349853283175619510255126908401163581602815539l9*t*u*v

[4]:
  _[1]=r-1.4
  _[2]=ew+0.620754943983586900941679142132077117923323153802Z4
  _[3]=ev+0.620754943983586900941679142132077117923323153802Z4
  _[4]=v
  _[5]=u^2-0.285063103849172883702919547894478304873393213308S4
  _[6]=t^2-0.285063103849172883702919547894478304873393213308S4
  _[7]=s
[5]:
  _[1]=r-1.4
  _[2]=ew+0.015665034671961754863974784720813826678603082481373
  _[3]=ev+0.627335517800769966061015627408554507173848648775G
  _[4]=v^2-2.032500495930121006949525695684269746961818258683GG
  _[5]=u
  _[6]=t^2-0.285063103849172883702919547894478304873393213308S4
  _[7]=s

There are numerical coefficients that are very lengthy ones. This is due to a problem in the algorithm in the Gröbner bases generation. The computational procedure applies the Buchberger's algorithm, in which the addition, subtraction, multiplication and division are iterated to the polynomial system. In the intermediate expression through the computation, some polynomials with huge degrees may arise, whose coefficients have extreme difference in the numerical scale. This difference in the scale of coefficients will remain in the final result [19]. To assure the numerical accuracy, we must resort to the computations with the arbitrary precision.

Though the solutions include complex ones, the physically meaningful real solutions are shown in Table. 1. We obtain four combinations, where the two electrons of up or down spins are located the symmetric or asymmetric wavefunctions. This means we obtain both of the ground and the excited states.

| | | | | |
|---|---|---|---|---|
| s (the coefficient for electron 1) | 0.00000 | -1.42566 | 0.00000 | -1.42566 |
| t (the coefficient for electron 1) | -0.53391 | 0.00000 | -0.53391 | 0.00000 |
| u (the coefficient for electron 2) | -0.53391 | -0.53391 | 0.00000 | 0.00000 |
| v (the coefficient for electron 2) | -0.53391 | 0.00000 | -1.42566 | -1.42566 |
| ev (the eigenvalue for electron 1) | -0.62075 | -0.01567 | -0.62734 | 0.01884 |
| ew (the eigenvalue for electron 2) | -0.62075 | -0.62734 | -0.01567 | 0.01884 |
| r (the interatomic distance) | 1.40000 | 1.40000 | 1.40000 | 1.40000 |
| Etot (The total energy) | -1.09624 | -0.49115 | -0.49115 | 0.15503 |
| electron1 | symmetric orbital | asymmetric orbital | symmetric orbital | asymmetric orbital |
| electron2 | symmetric orbital | symmetric orbital | asymmetric orbital | asymmetric orbital |

Table 1. This shows the solutions for equations (18). The electron 1 and 2 lie in the up- and down- spin respectively.

In order to obtain the ground state alone, we add (19) into the equations in (14).

$$s = v = 0 \tag{19}$$

(This is the trick applicable to this example only. In general, the ground state is given as a solution where the sum of the total occupied eigenvalue becomes minimum one. To specify the ground state, it is enough to compute eigenvalues alone. For this purpose, in making the triangular decomposition of the equations, we can prepare the equations including only eigenvalues as unknowns. We have only to solve them.) With this treatment, we can replace the equation to be solved with a simpler one. The interatomic distance r and the wavefunctions are optimized at the same time, as is done in Car-Parrinello method. The part of the equations including r is shown in (20), whose real solutions are shown in Table 2.

[1]: (20)
8942144364*r^23-435341589039*r^22+9813157241157*r^21-134458128500631*r^20+1251986164962728*r^19-8584760758387395*r^18+48176522279858253*r^17-254992901607817871*r^16+1360184656773665254*r^15-668541270541318423

5*r^14+26848712421674517351*r^13-82265960807423324641*r^12+185370480318135661708*r^11-295651827763150999108*r^10+307426892321213994312*r^9-148683667595876075980*r^8-97338526988608612178*r^7+245772518836579791529*r^6-200002425723099153061*r^5+47298638179277635737*r^4+46006348188804187952*r^3-41646082527529600720*r^2+13118922400543578496*r-1869747053688110592=0

[2]:
10313892*r^15-376866027*r^14+6245669754*r^13-61144647973*r^12+387764699571*r^11-1646957525797*r^10+4691411679124*r^9-8760215434992*r^8+10281598671237*r^7-7316755042677*r^6+3784010771997*r^5-2194016637700*r^4-299532295668*r^3+1482785614608*r^2-746000940352*r+36100845312=0

[3]:
10313892*r^15-376866027*r^14+6245669754*r^13-61144647973*r^12+378596795571*r^11-1405917509797*r^10+1917133055124*r^9+9044372533008*r^8-55118065080763*r^7+103464030245323*r^6+92432281739997*r^5-770797010005700*r^4+1063674493728332*r^3+652030557238608*r^2-2854269358708352*r+1954998898509312=0

|          | r      | ev      | t     |
|----------|--------|---------|-------|
| Solution1 | -1.812 | -6.6    | 0.846 |
| Solution2 | 1.652  | -0.578  | 0.545 |
| Soultion3 | 6.010  | -17.585 | 0.983 |

Table 2. This shows the real solutions in the equations (20).

The solutions include the positive and negative real valued ones and the imaginary valued ones. The meaningful solutions (r>0) are two in number. However, the solution which lies in the valid range of the Taylor expansion is only that of r∼1.6. The discrepancy between the solution and the experimental value ( r ∼1.4 ) is due to the numerical error caused by the roughness of the fourth order Taylor expansion and the rationalization of the numerical coefficients. In addition, it is also due to the not-optimized orbital exponent in the trial wavefunctions.

Next, the example of the solving of the inverse problem is demonstrated. Suppose a problem, where the energy difference between the occupied and the unoccupied states has a certain value: we should evaluate the interatomic distance r at which the energy difference shows this value. This example is a miniature of the inverse problem to find the lattice constants at which the band gap shows the desired breadth. In this case, we execute RHF calculations. The wavefunctions of the occupied and the unoccupied are given in (20) and (21). The eigenvalues are denoted as $e_{occ}$, $e_{unocc}$.

$$\phi_{occ}(x) = s(\exp(-x_A) + \exp(-x_B))/\sqrt{\pi} \qquad (21)$$

$$\phi_{unocc}(x) = t(\exp(-x_A) - \exp(-x_B))/\sqrt{\pi} \qquad (22)$$

The required equations are presented in (23), whose details are omitted here. The set of the equations for the occupied state can be obtained by the same way as was done as the example of UHF calculation. The orthogonality condition to the occupied and the unoccupied states is added to it.

$$\begin{aligned}
\frac{\partial \Omega}{\partial s} &= 0 \\
\frac{\partial \Omega}{\partial t} &= 0 \\
\frac{\partial \Omega}{\partial (e_{occ})} &= \langle \phi_{occ} | \phi_{occ} \rangle - 1 = 0 \\
\frac{\partial \Omega}{\partial (e_{unocc})} &= \langle \phi_{unocc} | \phi_{unocc} \rangle - 1 = 0 \\
\langle \phi_{occ} | \phi_{unocc} \rangle &= 0 \\
e_{occ} - e_{unocc} &= E_{gap}
\end{aligned} \qquad (23)$$

For example, let us compute r which gives $E_{gap} = e_{unocc} - e_{occ} = 0.9$. The real solutions are those at r=−1.103, 0.307, 1.643, 3.958. The solution in the valid range of the Taylor expansion is only that at r=1.643. The eigenvalues of the occupied and the unoccupied states are shown in Fig.3 as the function of R. It shows this result is the proper one.

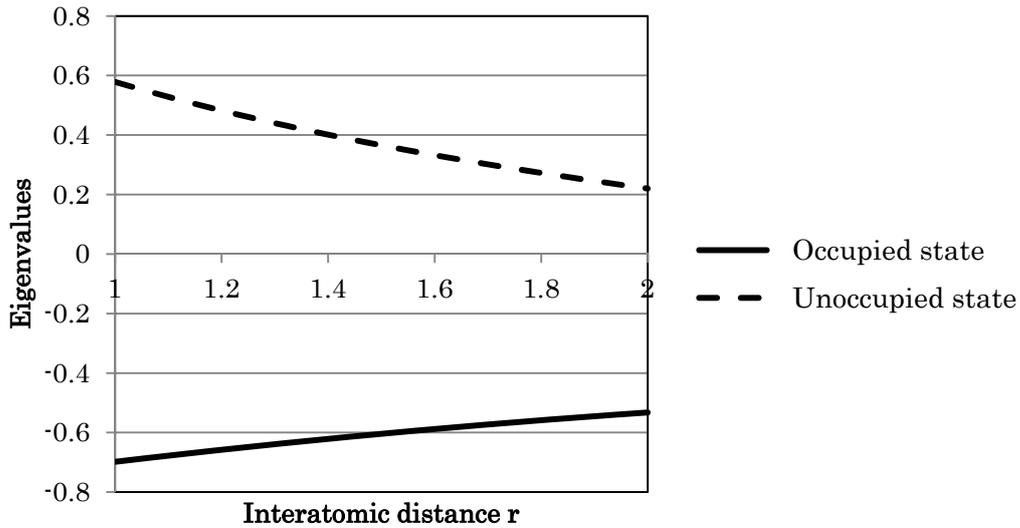

Figure 3. This figure shows the dependence of occupied and unoccupied eigenvalues on the interatomic distance. The eigenvalues of the occupied and the unoccupied states are shown by the real and broken lines respectively.

Then, in this case, is the structure is stable? If we evaluate the interatomic forces, it can be easily judged. On the other hand, with the view of symbolic-numeric solving, we can make use of the following judgment. To do this, in the set of the equations in (23), we insert the condition of eq. (24), the minimization condition of the energy functional with respect to the interatomic distance r.

$$\frac{\partial \Omega}{\partial r} = 0 \qquad (24)$$

The computed Gröbner base for it is {1} (as a set of polynomials). It includes only a constant polynomial "1". The zeros of the Gröbner bases provide us the solution of the equations. However, the term {1}, as a polynomial, does not become zero. Thus we can conclude that the supposed problem does not have the solution with the stable structure.

Next, we will show another numerical method by means of Stickelberger's theorem. The example is the optimization of interatomic distance in $H_2$, where the RHF calculation is executed. The trial wavefunction is given as (25).

$$\phi_{occ}(r) = t(\exp(-r_A) + \exp(-r_B))/\sqrt{\pi} \qquad (25)$$

The eigenvalue is denoted as ev and the interatomic distance is r. The HFR equation becomes the set of polynomials expressed by t, ev, and r. This set of equations, in the mathematical sense, constructs a zero-dimensional ideal in the polynomial ring, the zeros of which correspond to a residue ring R/I. (R means the rational number field.) A is a finite-dimensional vector space over R. Its base is represented by monomials of t, ev, and r, which are shown in (26).

| | | | | | |
|---|---|---|---|---|---|
| B[1]=t*ev*r^3 | b[2]=t*r^4 | b[3]=t^3*ev | b[4]=t*ev^3 | b[5]=t^3*r | (26) |
| b[6]=t*ev^2*r | b[7]=t*ev*r^2 | b[8]=t*r^3 | b[9]=ev*r^3 | b[10]=r^4 | |
| b[11]=t^3 | b[12]=t^2*ev | b[13]=t*ev^2 | b[14]=ev^3 | b[15]=t^2*r | |
| b[16]=t*ev*r | b[17]=ev^2*r | b[18]=t*r^2 | b[19]=ev*r^2 | b[20]=r^3 | |
| b[21]=t^2 | b[22]=t*ev | b[23]=ev^2 | b[24]=t*r | b[25]=ev*r | |
| b[26]=r^2 | b[27]=t | b[28]=ev | b[29]=r | b[30]=1 | |

The transformation matrix corresponding to the multiplication by the variable t is shown in (27).

$$
\begin{pmatrix}
0. & 0. & 0. & 0. & 0. & 0. & 0. & 0. & 1. & 0. & 0. & 0. & 0. & 0. & 0. & 0. & 0. & 0. & 0. \\
0. & 0. & 0. & 0. & 0. & 0. & 0. & 0. & 0. & 1. & 0. & 0. & 0. & 0. & 0. & 0. & 0. & 0. & 0. \\
0. & 0. & 0. & 0. & 0. & 0. & 0. & 0. & 0. & 0. & 0. & 1. & 0. & 0. & 0. & 0. & 0. & 0. & 0. \\
0. & 0. & 0. & 0. & 0. & 0. & 0. & 0. & 0. & 0. & 0. & 0. & 0. & 1. & 0. & 0. & 0. & 0. & 0. \\
0. & 0. & 0. & 0. & 0. & 0. & 0. & 0. & 0. & 0. & 0. & 0. & 0. & 0. & 0. & 1. & 0. & 0. & 0. \\
0. & 0. & 0. & 0. & 0. & 0. & 0. & 0. & 0. & 0. & 0. & 0. & 0. & 0. & 0. & 0. & 0. & 0. & 1. \\
2.46857 & 0.250358 & 0.000433932 & -0.519297 & 0.000905017 & -0.229989 & 0.400711 & 0.0592776 & 0. & 0. & 0.00037686 & 0. & -0.0488928 & 0. & 0. & 0.0253457 & 0. & 0.0106224 & 0. \\
2.85865 & 1.8302 & 0.010766 & -3.25489 & 0.00562066 & 0.195796 & 0.755874 & 0.513375 & 0. & 0. & -0.0000321824 & 0. & 0.188933 & 0. & 0. & 0.177659 & 0. & 0.101971 & 0. \\
0. & 0. & 0. & 0. & 0. & 0. & 0. & 0. & 0. & 0. & 0. & 0. & 0. & 0. & 0. & 0. & 0. & 0. & 0. \\
-3.28726 & 16.9846 & 0.815632 & -68.3604 & 0.199106 & -6.99935 & 17.2255 & 12.5718 & 0. & 0. & 0.010156 & 0. & 2.94616 & 0. & 0. & 8.19509 & 0. & 3.31843 & 0. \\
0. & 0. & 0. & 0. & 0. & 0. & 0. & 0. & 0. & 0. & 0. & 0. & 0. & 0. & 0. & 0. & 0. & 0. & 0. \\
1.10471 & 0.0541875 & 0.0005962 & -0.43266 & 0.000241163 & 0.0161693 & 0.169585 & 0.0191714 & 0. & 0. & 4.36536 \times 10^{-6} & 0. & 0.00658837 & 0. & 0. & 0.0121792 & 0. & 0.00422577 & 0. \\
-41.1091 & 92.4474 & 0.457252 & -248.171 & 0.949199 & 7.93844 & -9.06291 & 33.1339 & 0. & 0. & 0.0550604 & 0. & 7.29159 & 0. & 0. & -1.91674 & 0. & 7.47555 & 0. \\
0. & 0. & 0. & 0. & 0. & 0. & 0. & 0. & 0. & 0. & 0. & 0. & 0. & 0. & 0. & 0. & 0. & 0. & 0. \\
0.381952 & -0.482168 & -0.00366255 & 1.05379 & -0.00214058 & -0.266467 & 0.0658373 & -0.170307 & 0. & 0. & -0.0000384651 & 0. & -0.0484318 & 0. & 0. & 0.00667468 & 0. & -0.0375156 & 0. \\
0. & 0. & 0. & 0. & 0. & 0. & 0. & 0. & 0. & 0. & 0. & 0. & 0. & 0. & 0. & 0. & 0. & 0. & 0. \\
-13.5886 & -1.50329 & 0.000693225 & 2.88068 & -0.00965266 & 1.64897 & -2.25783 & -0.319754 & 0. & 0. & -0.00357141 & 0. & 0.303898 & 0. & 0. & -0.125453 & 0. & -0.05279 & 0. \\
-9.85681 & -9.15049 & -0.0964526 & 22.4732 & -0.0359414 & -1.57568 & -3.20362 & -3.05264 & 0. & 0. & 0.00102063 & 0. & -1.41532 & 0. & 0. & -1.22295 & 0. & -0.657454 & 0. \\
480.008 & -277.275 & -3.39938 & 331.95 & 0.518864 & -172.803 & 89.1965 & -51.8746 & 0. & 0. & 0.670983 & 0. & -36.5975 & 0. & 0. & 14.0368 & 0. & 1.13365 & 0. \\
0. & 0. & 0. & 0. & 0. & 0. & 0. & 0. & 0. & 0. & 0. & 0. & 0. & 0. & 0. & 0. & 0. & 0. & 0. \\
7.34299 & 6.6392 & 0.00938662 & -17.0623 & 0.0147453 & 1.43466 & 1.18689 & 1.56243 & 0. & 0. & -0.000518309 & 0. & 0.363976 & 0. & 0. & 0.111898 & 0. & 0.278794 & 0. \\
0. & 0. & 0. & 0. & 0. & 0. & 0. & 0. & 0. & 0. & 0. & 0. & 0. & 0. & 0. & 0. & 0. & 0. & 0. \\
15.2609 & -1.66537 & -0.0325943 & 4.78209 & -0.0255023 & -8.07839 & 2.60648 & -0.91568 & 0. & 0. & 0.00589181 & 0. & -0.519365 & 0. & 0. & 0.0343565 & 0. & -0.229064 & 0. \\
4.37903 & 18.2502 & 0.31662 & -48.1151 & 0.0770211 & 4.75029 & 3.82373 & 7.43136 & 0. & 0. & -0.0068586 & 0. & 3.54526 & 0. & 0. & 3.3935 & 0. & 1.7193 & 0. \\
0. & 0. & 0. & 0. & 0. & 0. & 0. & 0. & 0. & 0. & 0. & 0. & 0. & 0. & 0. & 0. & 0. & 0. & 0. \\
176.488 & 32.4122 & 0.0776154 & -62.9606 & 0.177886 & -12.346 & 26.5689 & 7.04326 & 0. & 0. & -0.0170918 & 0. & -5.2778 & 0. & 0. & 1.56207 & 0. & 1.18349 & 0. \\
-21.5227 & -39.0633 & -0.494407 & 121.758 & -0.355958 & 0.576694 & -5.67678 & -17.58 & 0. & 0. & -0.00940279 & 0. & -3.59933 & 0. & 0. & -4.92509 & 0. & -4.18857 & 0. \\
22.4313 & 97.9958 & 1.19706 & -159.224 & 0.120322 & 24.8872 & 3.17625 & 23.7835 & 0. & 0. & -0.11519 & 0. & 5.8433 & 0. & 0. & 1.73174 & 0. & 2.14857 & 0.
\end{pmatrix}
$$
(27)

The other transformation matrices are calculated in similar ways. We can evaluate the eigenvector $v_\xi$ of the matrix $m_t$ and can obtain eigenvalues corresponding to variables t and ev by means of the relations in (28) and (29). The real valued solutions are identical to those in Fig.9, given as the previous example.

$$m_t \cdot v_\xi = \xi_t \cdot v_\xi \qquad (28)$$

$$m_{ev} \cdot v_\xi = \xi_{ev} \cdot v_\xi \qquad (29)$$

## 4. Summary and discussion

The advantages by the present method are recapitulated here.

(I)In the present method, the fundamental equation (the HFR equation and the constraints) is expressed as a set of polynomial equations, in which the electronic structure calculations and the extra additional conditions, are combined with each other seamlessly. In the conventional method, the loop for eigenvalue solutions, that of SCF (Self Consistent Field) procedure, and that of the optimization calculation are nested with one another, being somewhat complicated. By contrast, in the present method, those nested loops are unified in a flow of the search for roots in a set of polynomial equations, which afford us clear view in the numerical computation and the shortcut to the needed information. It is not necessary to iterate the independent optimization to the atomic structure and the wavefunctions, as is done in the conventional method. The present method will simulate the cases where the dynamics of the wavefunctions and that of the nuclei are strongly coupled with each other, beyond the adiabatic approximation.

(II) The merit in expressing the basic HFR equation and the constraints by means of the set of polynomial equations is as follows. The polynomial sets inform us the relationship among unknown variables. Thus, in order to evaluate those unknowns, we can divide suitable parts of them into the prepared inputs and the expected outputs, respectively. The calculation is not confined to the conventional framework, such as, whose the input is the structure and the output is the electronic the states. The distinction between the forward and the inverse problems are cleared away, and we can treat all of them as the forward problems in a unified way. In order to cope with the inverse problem, we should check whether it is well-posed or not. The present method affords us a key to this. After the transformation from the fundamental equation to the Gröbner bases, the present method can inspect the properness of the problem, i.e. the existence of the solutions. Based on the mathematical ideal theory, it can judge the existence of solution which provides us the zeros of the Gröbner bases. If the solutions exist, it can also determine whether these are isolated points or the sets with the more than one dimension. If the solutions are isolated, the numbers of the solutions, in the range of the real or complex numbers, can be known. (See references.)

(III) The merit of the present method applied to SCF calculations is as follows. The roots of the triangulated equations are obtained one by one numerically. The eigenvalue solutions for multi-dimensional matrix and the iterative approximation for the mean field, where the difficulty in the convergence generally occurs, are unnecessary.

(IV) This method has the following merit in the calculation for the optimization. For example, consider the optimization of the atomic structure. By means of the elimination of the variables, we can obtain a set of equations which have only atomic coordinates. The roots are the stable optimized structure. It means that the "pseudo" atomic interactions are obtained without SCF calculations. In the optimization of other parameters, there are similar merits.

The present work shows that the concept of the "molecular orbital algebraic equation" by means of the "polynomial approximation to molecular integrals" is applicable to the realistic first principles electronic structure calculation, as well as its potentiality to several fundamental problems which are difficult to be handled by the conventional method. In a pity, at present, this study does not necessarily afford us sufficiently precise calculation. One reason to this is the constraint by the ability of the hardware. The cost in the symbolic computation is so massive that we are obliged to reduce the degrees of the approximating polynomial and rationalize numerical coefficients in lower accuracies. There is also a fundamental problem in the theoretical side. As can be seen in the starting polynomial equation of (14), the scales of the numerical coefficients are almost similar. While, after the symbolic computations, in the generated Gröbner bases, the scales of the coefficients show great discrepancies. Although the extreme growth of coefficients as this results in the larger computational costs and the lowering of the accuracies in the numerical procedure, highly effective remedies are few at now. However, the improvement in the computer architecture is so rapid that we can expect the achievement of the sufficient accuracy by the present method and its application to complex and large material in future, aided by the refinement of the symbolic computation theory.

**Appendix. On the concept of algebraic molecular orbital equations**

The concept of algebraic molecular orbital equations [6-9] is outlined in this appendix. We express molecular orbitals by the linear combination of atomic orbitals (LCAO).

$$\phi_k = \sum_\alpha^{atom} \sum_i^{base} C_{ai}^k \chi(R_\alpha, \{n_i, l_i, m_i\}, \zeta_i, r, \theta, \phi) \quad (A.1)$$

The key components to the molecular orbital calculations are molecular integrals, which are the matrix elements of the each part of the Hamiltonian operator, such as, kinetic, nuclear and electronic potential, and overlapping elements, obtained by the use of LCAO. The molecular integrals can be expressed as multi-variable analytic functions. Since there is some difficulty in mathematical operations on multi-variable analytic functions, we will approximate them as multi-variable polynomials. The approximation to molecular integrals is obtained by Taylor expansion;

$$f(x) = \sum_{p=0}^\infty \frac{1}{p!} \sum_{k=0}^p \binom{p}{k} (x - x_0)^{p-k} f(x_0)^{(p)} x^p \cong \sum_{i=0}^N A_i(x_0) x^i \quad (A.2)$$

As an example of the molecular intergrals, the two-centered overlapping integrals is defined in (A.3)

$$S_{AB}^{ab} \equiv \int \chi(R_A, \{n_a, l_a, m_a\}, \zeta_a, r_A, \theta_A, \phi_A) \chi(R_B, \{n_b, l_b, m_b\}, \zeta_b, r_B, \theta_B, \phi_B) dr^3. \quad (A.3)$$

The integration generates an analytic function of two orbital exponents and inter-atomic distance R. The approximation is given as

$$S_{AB}^{ab}(\zeta_a, \zeta_b, R) \cong \sum_{p_a, p_b, p_R} A(\{n_a, l_a, m_a\}_A, \{n_b, l_b, m_b\}_B)_{p_a, p_b, p_c} \zeta_a^{p_a} \zeta_b^{p_b} R^{p_R} \quad (A.4)$$

The other molecular integrals can be expressed in the similar way. The Hartree-Fock-Roothann equation and the energy functional with constraints are constructed from those integrals. Once the molecular integrals are approximated as polynomials, the Hartree-Fock-Roothann equation ($H[\Psi] \cdot \Psi = S \cdot \psi \cdot \epsilon$) and the energy functional $\Omega[\{\Psi, \epsilon\}]$ take polynomial expressions. The orbital exponents and atomic coordinates can equivalently be regarded as parameters in the calculus of variations, as well as LCAO coefficients. It will be adequate to call this multi-variable polynomial expression "algebraic molecular orbital equation." It is not necessary to regard those equations as pure numerical eigenvalue problems. Those are a set of polynomials, to which both symbolic manipulations and numerical solving are applicable.

## Acknowledgements

In this research, the analytic formulas and the Taylor expansions of molecular integrals were generated by Symbolic computation software Mathematica. The symbolic-numeric solving was executed by Computer algebra system Singular.

The author wishes to acknowledge his colleague Dr. J. Yasui for discussions on the concept of molecular orbital algebraic equations. The author also thanks Dr. Yasui for providing him with the symbolic computation software to generate STO molecular integrals.

The author wishes to gratefully acknowledge Prof. Toshiko Kikuchi (菊地敏子) and Dr. Ichio Kikuchi (菊地市夫) for their guide to higher algebra and geometry and thanks for their kind advices.